\documentclass[10pt,twocolumn,oneside,final]{IEEEtran}

\usepackage{times, amsfonts}
\usepackage[cmex10]{amsmath}
\usepackage{array}
\usepackage{algorithm,algcompatible}
\usepackage[font=footnotesize]{subfig}

\usepackage{color}
\usepackage{xcolor}
\usepackage{multirow}
\usepackage{url,cite}
\usepackage{times}
\usepackage{amsmath}
\usepackage{amssymb}
\usepackage{mathrsfs}
\usepackage{amsfonts}
\usepackage{graphicx}
\usepackage{bm}
\usepackage{amsthm}
\usepackage{booktabs}
\usepackage{tabularx}

\algnewcommand\INPUT{\item[\textbf{Input:}]}%
\algnewcommand\OUTPUT{\item[\textbf{Output:}]}%

\newcommand{\E}{{\mathbb{E}}} 
\newcommand{\Var}{{\mathbb{V}\mathrm{ar}}} 
\newcommand{\CN}{{\mathcal{CN}}} 
\newcommand{\C}{{\mathbb{C}}} 
\newcommand{\tr}{{\mathrm{tr}}} 
\newcommand{\Hm}{{\mathrm{H}}} 
\newcommand{\T}{{\mathrm{T}}} 
\newcommand{\aseq}{{\overset{\mathrm{a.s.}}{=}}} 

\theoremstyle{plain}
\newtheorem{theorem}{Theorem}
\newtheorem{lemma}{Lemma}
\newtheorem{proposition}{Proposition}
\newtheorem{corollary}{Corollary}

\theoremstyle{definition}

\theoremstyle{remark}
\newtheorem{remark}{\bf{Remark}}

\begin{document}

\title{Multipair Massive MIMO Relaying with Pilot-Data Transmission Overlay}
\author{Leyuan Pan, \IEEEmembership{Student Member, IEEE,}
    \and Yongyu Dai, \IEEEmembership{Student Member, IEEE,}
    \and Wei Xu, \IEEEmembership{Senior Member, IEEE,}\\
and \and Xiaodai Dong, \IEEEmembership{Senior Member, IEEE}}

\maketitle

\begin{abstract}
	We propose a pilot-data transmission overlay scheme for multipair massive multiple-input multiple-output (MIMO) relaying systems employing either half- or full-duplex (HD or FD) communications at the relay station (RS).
	In the proposed scheme, pilots are transmitted in partial overlap with data to decrease the channel estimation overhead.
	The RS can detect the source data with minimal destination pilot interference by exploiting the asymptotic orthogonality of massive MIMO channels.
	Then pilot-data interference can be effectively suppressed with assistance of the detected source data in the destination channel estimation.
	Due to the transmission overlay, the effective data period is extended, hence improving system throughput.
	Both theoretical and simulation results confirm that the proposed pilot-data overlay scheme outperforms the conventional separate pilot-data design in the limited coherence time interval scenario.
	Moreover, asymptotic analyses at high and low SNR regions demonstrate the superiority of the proposed scheme regardless of the coherence interval length.
	Because of simultaneous transmission, the proper allocation of source data transmission and relay data forwarding power can further improve the system performance.
	Hence a power allocation problem is formulated and a successive convex approximation approach is proposed to solve the non-convex optimization problem with the FD pilot-data transmission overlay.
\end{abstract}


\IEEEpeerreviewmaketitle

\section{Introduction}\label{sec:Introduction}
\IEEEPARstart{M}{assive} multiple-input multiple-output (MIMO) technology is becoming one of the most promising solutions for the next-generation wireless communication to meet the urgent demands of high-speed data transmissions and explosive growing numbers of user terminals, such as the traditional mobile equipments and the new Internet of Things (IoT) devices \cite{Marzetta2010,Hieving2013,Hoydis2013,Larsson2014Massive,Chin2014}.
Compared with conventional MIMO mechanisms, massive MIMO is capable of achieving higher reliabilities, increased throughputs and improved energy efficiency by employing less complicated signal processing techniques, e.g., maximum-ratio combining/maximum-ratio transmission (MRC/MRT), with inexpensive and low-power components \cite{Larsson2014Massive}.
Despite of the advantages, massive MIMO is also facing significant challenges on the way towards practical applications.
How to obtain precise channel state information (CSI) while consuming limited resources is most critical and fundamental.

In existing massive MIMO studies, time-division duplex (TDD) is more widely considered than frequency-division duplex (FDD) because it is potentially easier and more feasible to obtain CSI \cite{Marzetta2006}.
Thanks to the channel reciprocity, a TDD system exploits uplink pilot training to estimate channels which can be used in both uplink and downlink data transmissions within a coherence interval where the interval length is determined by the mobility of user equipment.
Compared with the downlink one, the uplink pilot training saves a large amount of resources to estimate the channels of a large-scale antenna array because each pilot sequence can be used to estimate the channels between all base station antennas and a single-antenna user equipment.
However, the uplink channel estimation has to deal with pilot contamination issues as the user number grows.
It is reported in \cite{Jose2011,Hieving2013} that pilot contamination reduces the system performance but cannot be suppressed by increasing the number of antennas.
Generally, the length of pilot sequence should be equal to or greater than the number of users to guarantee the orthogonality of pilot patterns among different users, in order to avoid pilot contamination.
In a massive MIMO system, due to the large user number, orthogonal pilot sequences become very long, causing significant overhead for channel estimation and thus degrading the effective system throughput.
When the channel varies with time due to medium to high mobility, i.e., relatively short coherence interval, the pilot overhead issue gets more severe as channel estimation needs to be done frequently.
There are some efforts in the literature to reduce pilot overhead in the massive MIMO cellular system serving a large number of users within a finitely long coherence interval.
Zhang \emph{et al.} proposed a semi-orthogonal pilot design in \cite{Zhang2014} and \cite{Zheng2014} to transmit both data and pilots simultaneously where a successive interference cancellation (SIC) method was employed to reduce the contaminations of interfering pilots.
You \emph{et al.} investigated the performance of a pilot reuse scheme in the single-cell scenario which distinguishes users by the angle of arrival and thereby reuses pilot patterns \cite{You2015}.
A time-shifted pilot based scheme was proposed in \cite{Fernandes2013} and it was then extended to the finite antenna regime in \cite{Jin2014} to cope with the multi-cell scenario.
Nevertheless, all the research introduced above focused on point-to-point (P2P) communications.
Few work has studied the pilot scheme design and optimization in massive MIMO relaying systems, especially for one-way multipair communications.

The relaying technique is an emerging cooperative technology capable of scaling up the system performance by orders of magnitude, extending the coverage and reducing power consumption\cite{DohlerLi2010}.
Combining with massive MIMO technologies whereby the relay station (RS) is equipped with large scale antenna arrays, the performance of a relaying system can be dramatically improved \cite{Ngo2013,NgoLarssonMemberEtAl2013,Suraweera2013,Dai2015,Ngo2014,Ngo2014a}.
Moreover, in spite of the conventional half-duplex (HD) system, the full-duplex (FD) relaying technique has attracted more interests recently due to the simultaneous uplink (multi-access of the sources to RS) and downlink (broadcasting of RS to the destinations) data transmissions, whereby the overall system performance is further improved \cite{Ngo2014,Ngo2014a,Rodriguez2014}.
However, similar to the P2P system, the multiuser relaying system also suffers from the critical channel estimation overhead within limited coherence time intervals.
For a relaying system, it may be even worse as both source and destination users need to transmit pilots within the coherence interval, where the coherence interval determined by user pairs may be shorter than or at best equal to that by each user.
Further, different from the P2P cellular system, the throughput of the whole relaying system is determined by the weaker one between the uplink and downlink connections. Thus, it is critical to co-consider both uplink and downlink transmissions when designing the pilot scheme for the relaying system, while previous work in the literature did not take this into consideration.

This paper investigates the pilot and data transmission scheme in multipair massive MIMO relaying systems for both HD and FD communications.
Due to the massive antennas equipped on RS, the source-relay and relay-destination channels are asymptotically orthogonal to each other, and thereby the transmission phase of pilots and data can be shifted to overlap each other to reduce the overhead of pilot transmission and accordingly to improve the system performance.
Based on this consideration, a transmission scheme with pilot-data overlay in both HD and FD communications is proposed in this paper.
Here, the main contributions of the paper are summarized as follows:
\begin{itemize}
	\item \textbf{Pilot-data overlay transmission scheme design:}
	A transmission scheme with pilot-data overlay in both HD and FD multipair massive MIMO relaying systems is proposed and designed.
	In the HD overlay scheme, destination pilots are transmitted simultaneously with source data transmission, such that the effective data transmission duration is increased.
	Moreover, both source and destination pilots are transmitted along with data transmission in the FD system and thus the effective data transmission duration can be further increased.
	However, pilot and data contaminate each other at the RS due to the simultaneous transmission.
	Nevertheless, by exploiting the asymptotic orthogonality of massive MIMO channels, this paper demonstrates that the received data and pilots can be well separated from each other with only residues of additive thermal noise by applying the MRC processing.
	After all, the effective data transmission duration is extended within the limited coherence time interval and therefore the overall system performance is improved.
	
	\item \textbf{Closed-form achievable rates and comparison with conventional schemes:}
	This paper derives \emph{closed-form expressions} of the ergodic achievable rates of the considered relaying systems with the proposed scheme.
	The derived expression reveals that the loop interference (LI) in the FD overlay scheme can be effectively suppressed by the growing number of RS antennas and no error propagation exists with the proposed scheme, which is a critical issue in \cite{Zhang2014} where a semi-orthogonal pilot design is applied to the P2P system.
	Numerical results show that the superiority of the proposed scheme persists even with 25 dB LI.
	For quantitative comparison between the proposed scheme and conventional ones, asymptotic achievable rates at both ultra-high and -low SNRs are derived and the superiority of the proposed scheme is proved theoretically.
	
	\item \textbf{Power allocation design:}
	This paper designs an optimal power allocation for the FD overlay scheme to minimize the interference between pilot and data transmissions by properly regulating the source and relay data transmission power for a fixed pilot power and proposes a successive convex approximation (SCA) approach to solve the non-convex optimization problem.
	Simulation results indicate that the proposed approach further improves the achievable rate compared with the equal power allocation.
	In addition, the proposed approach is computationally efficient and converges fast.
	With typical configurations (e.g., total data transmission power at 20 dB), simulation shows that the proposed approach converges to a relative error tolerance at $\epsilon=10^{-5}$ after a few, say 4, iterations.
\end{itemize}

\emph{Organization:}
The rest of this paper is organized as follows.
The channel and signal models are presented in Section \ref{sec:SystemModel} within which the conventional and overlay scheme is presented and proposed, respectively.
In the following Section \ref{sec:ChannelEstimation}, channel estimations of the proposed scheme applying to both HD and FD relaying systems are elaborated in details and the system achievable rates are derived theoretically in Section \ref{sec:AchievableRateAnalysis}.
Section \ref{sec:AsymptoticAnalysis} and \ref{sec:PowerAllocation} extend the analyses to the asymptotic scenario and power allocation consideration, respectively.
The results presented in Section \ref{sec:NumericalResults} reveal the performance comparisons numerically.
Section \ref{sec:Conclusion} concludes our works.

\emph{Notations:}
The boldface capital and lowercase letters are used to denote matrices and vectors, respectively, while the plain lowercase letters are scalars.
The superscript $(\cdot)^*$, $(\cdot)^\T$ and $(\cdot)^\Hm$ stands for the conjugate, transpose and conjugate-transpose of a vector or matrix, respectively.
$\mathbf{I}_M$ represents the identity matrix of size $M$.
The operator $\|\cdot\|$, $\|\cdot\|_\mathrm{F}$ and $\mathrm{tr}(\cdot)$ denotes the Euclidean norm of a vector, the Frobenius norm and the trace of a matrix, respectively.
For statistical vectors and matrices, $\E\{\cdot\}$ and $\Var\{\cdot\}$ are utilized to represent the expectation and variance, respectively.
Moreover, $\aseq$ is used to denote the almost sure convergence and the notation $\mathbf{x}\sim\CN(\mathbf{0}, \bm{\Sigma})$ represents the complex Gaussian random vector $\mathbf{x}$ with zero mean and covariance matrix $\bm{\Sigma}$.
Finally, the postfix $[\iota]$ denotes the corresponding signal in the $\iota$th coherence interval.

\section{System Model}\label{sec:SystemModel}
\subsection{Signal and Channel Model}\label{sec:SignalChannelModel}
\begin{figure}[htbp]
	\centering
	\includegraphics[width=\linewidth]{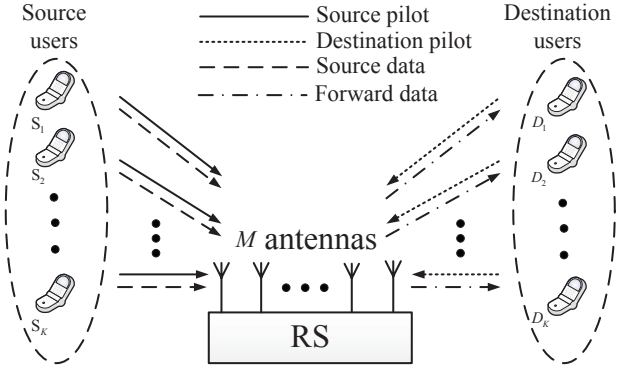}
	\caption{System diagram.}
	\label{fig:SystemModel}
\end{figure}
As depicted in Fig. \ref{fig:SystemModel}, this paper considers both HD and FD one-way relaying systems where $K$ pairs of single-antenna source and destination users are served by RS equipped with $M$ ($M\gg K\gg 1$) antennas.
In the following sections, the noise power is normalized to 1.
Let $\rho_\mathrm{p}$, $\rho_\mathrm{s}$ and $\rho_\mathrm{d}$ be the transmission power of pilots, source and forward data, respectively.
The channel matrices from sources and destinations to RS are denoted by $\mathbf{G}_\mathrm{s}\in\C^{M\times K}$ and $\mathbf{G}_\mathrm{d}\in\C^{M\times K}$, which are concisely named as source and destination channels, respectively, where the $k\mathrm{th}$ column of either matrix, $\mathbf{g}_{\mathrm{s}k}$ or $\mathbf{g}_{\mathrm{d}k}$, stands for the channel vector from the $k\mathrm{th}$ corresponding user to RS. 
Both channel matrices are decomposed as $\mathbf{G}_\mathrm{s}=\mathbf{H}_\mathrm{s}\mathbf{D}_\mathrm{s}^{1/2}$ and $\mathbf{G}_\mathrm{d}=\mathbf{H}_\mathrm{d}\mathbf{D}_\mathrm{d}^{1/2}$, where the large-scale fading matrices $\mathbf{D}_\mathrm{s}$ and $\mathbf{D}_\mathrm{d}$ are both diagonal with the $k\mathrm{th}$ entries as $\beta_{\mathrm{s}k}$ and $\beta_{\mathrm{d}k}$, respectively, and the small-scale fading matrices $\mathbf{H}_\mathrm{s}$ and $\mathbf{H}_\mathrm{d}$ are constructed by independent identically distributed (i.i.d.) $\CN(0, 1)$ random variables (RVs).
The leakage channel of the FD communication is modeled as $\mathbf{G}_\mathrm{LI}\in\C^{M\times M}$ which consists of the large-scale fading factor denoted by $\beta_\mathrm{LI}\mathbf{I}_M$ and the small-scale fading matrix constructed by i.i.d. $\CN(0, 1)$ RVs, where $\beta_\mathrm{LI}$ is the leakage power gain after LI cancellations \cite{Riihonen2011}.
According to the experiments on MIMO FD self-interference cancellation presented in \cite{Bharadia2014}, the received leakage power can be suppressed to a level almost comparable to the noise floor regardless of the transmission power (see Fig. 10 of \cite{Bharadia2014}).
Therefore, the average LI power, $\rho_\mathrm{LI}=\rho_\mathrm{d}\beta_\mathrm{LI}$, is assumed to be a constant in the following analysis.
Further, it is assumed that the same frequency band is reused for both uplink and downlink transmissions, and they obey the reciprocity, i.e., the channel matrices are consistent within a coherence time interval for both uplink and downlink communications.

\begin{figure*}[htbp]
	\centering
	\subfloat[HD conventional pilot-data transmission scheme]{
		\label{fig:PhaseDiagramHalfDuplexOrthogonal}
		\begin{minipage}[t]{0.48\linewidth}
			\centering
			\includegraphics[width=0.8\linewidth]{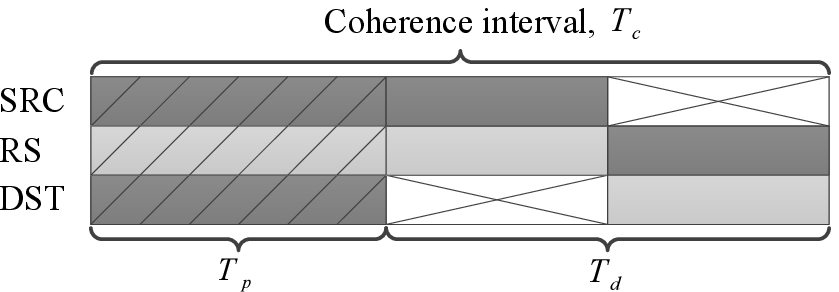}
		\end{minipage}
	}
	\hfill
	\subfloat[FD conventional pilot-data transmission scheme]{
		\label{fig:PhaseDiagramFullDuplexOrthogonal}
		\begin{minipage}[t]{0.48\linewidth}
			\centering
			\includegraphics[width=0.8\linewidth]{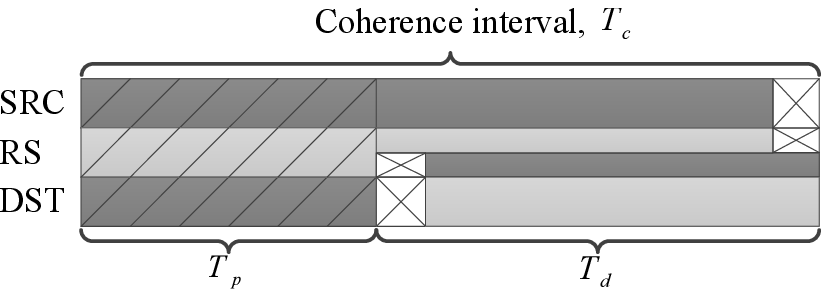}
		\end{minipage}
	}
	
	\subfloat[HD pilot-data overlay transmission scheme]{
		\label{fig:PhaseDiagramHalfDuplexSemiOrthogonal}
		\begin{minipage}[t]{0.48\linewidth}
			\flushleft
			\includegraphics[width=\linewidth]{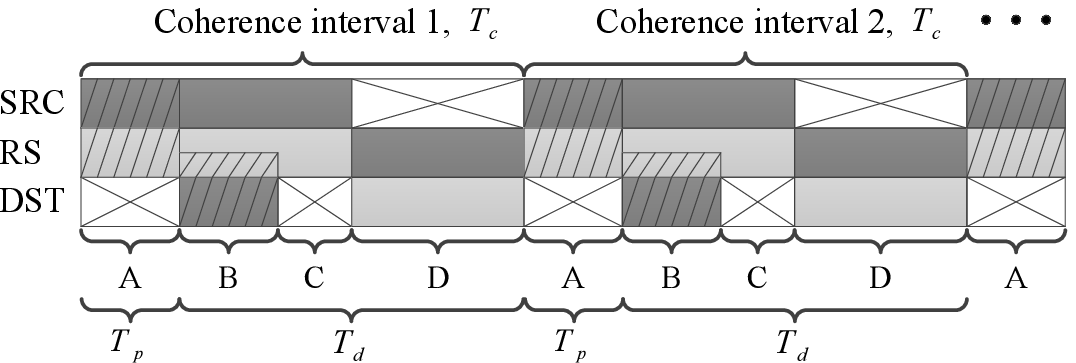}
		\end{minipage}
	}
	\hfill
	\subfloat[FD pilot-data overlay transmission scheme]{
		\label{fig:PhaseDiagramFullDuplexSemiOrthogonal}
		\begin{minipage}[t]{0.48\linewidth}
			\flushleft
			\includegraphics[width=\linewidth]{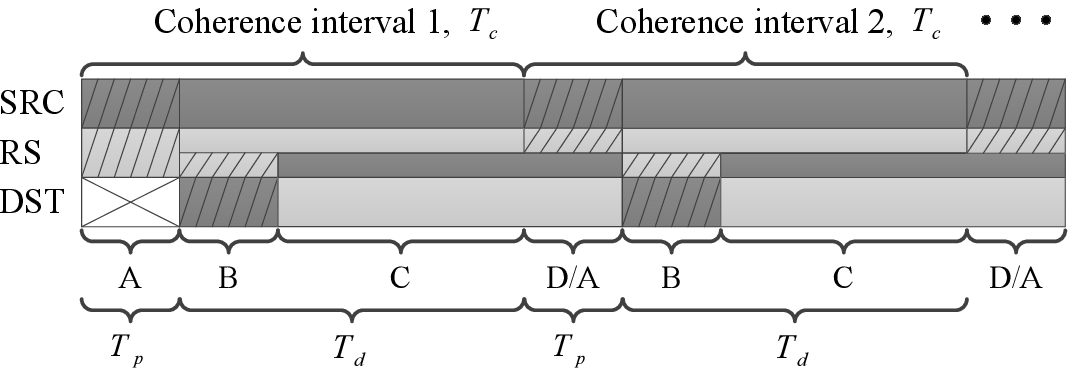}
		\end{minipage}
	}
	\vspace{5pt}
	\subfloat{
		\begin{minipage}[t]{0.8\linewidth}
			\centering
			\includegraphics[width=\linewidth]{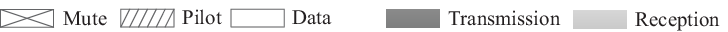}
		\end{minipage}
	}
	\caption{Conventional and proposed pilot-data transmission diagrams in both HD and FD one-way relaying systems, where $T_c$, $T_p$ and $T_d$ denote the lengths of coherence, pilot and data transmission intervals, respectively. (SRC: source users, RS: relay station, DST: destination users.)}
	\label{fig:PhaseDiagram}
\end{figure*}

\subsection{Conventional Pilot-data Transmission Scheme}
This subsection considers the conventional pilot-data transmission scheme for both HD and FD communications.

With regard to the HD system shown by Fig. \ref{fig:PhaseDiagram}(a), TDD is selected as the working mode where the source and destination users firstly transmit training pilots to help RS estimate CSI, following which the sources send uplink data to RS, and then the data is forwarded to destination users \cite{Marzetta2010,Fernandes2013,cui2014multi}.
Similarly, the FD system depicted in Fig. \ref{fig:PhaseDiagram}(b) shows that the CSI is also estimated first and the data is transmitted subsequently.
Yet the FD RS forwards data to destinations simultaneously with source data transmission by some negligible processing time delay\cite{Riihonen2011}.

In conventional MIMO systems, the pilot scheme is always designed by utilizing orthogonal pilot sequences to prevent the inner-cell pilot contamination which requires the length of pilot sequences to be not smaller than the number of users.
Therefore, $T_p\geq 2K$, where $T_p$ is the length of orthogonal pilot sequences.
The overhead of channel estimation is at least $2K/T_\mathrm{c}$ of each terminal, where $T_\mathrm{c}$ is the length of coherence time interval in terms of the number of symbol duration\cite{Ngo2014,Ngo2014a}.
In the massive MIMO relaying system, this overhead is extremely large because the number of users becomes large with the increasing antenna number while the coherence interval is, to some extent, fixed mainly depending on the mobility of terminals. Therefore, especially for the massive MIMO with a relatively short $T_\mathrm{c}$, most of the effective resource would be occupied by pilots, which makes the massive MIMO transmission inefficient. In order to reduce the pilot overhead and accordingly improve the data transmission efficiency, this paper proposes a pilot-data transmission overlay scheme in the following subsection.

\subsection{Pilot-data Overlay Transmission Scheme}
This paper proposes a pilot-data overlay transmission scheme for both HD and FD one-way relaying systems.
The general design of the proposed scheme is explained in this subsection while the detailed signal transmissions will be described mathematically in Sections \ref{sec:ChannelEstimation} and \ref{sec:AchievableRateAnalysis}.

With respect to the HD relaying system, the pilot-data overlay is depicted in Fig. \ref{fig:PhaseDiagram}(c).
To concisely describe signal transmissions, we separate a coherence interval into four phases denoted by A, B, C and D.
During phases A and B, pilots are transmitted from users to RS, while in C and D, the sources and RS conduct data transmissions, respectively.
Within phase A, all source users send piece-wisely orthogonal pilot sequences to assist RS in estimating channels while destination users keep mute, thus source channels can be estimated at RS without being contaminated.
Subsequently, destination users start transmitting pilots in phase B, while sources can send uplink data to RS simultaneously.
The RS observes both source data and destination pilots in this phase.
With the source channel estimated in phase A and the quasi-orthogonality of source and destination massive MIMO channels, the RS detects the source data and then is able to cancel it from the received signal, from which obtains the estimates of the destination channels.
Thereafter, sources keep sending uplink data in phase C and RS forwards downlink data to destination users in phase D.
In the following coherence intervals, the HD relaying system repeats these communication procedures.

Regarding the FD pilot-data overlay scheme, the communication procedure is shown in Fig. \ref{fig:PhaseDiagram}(d).
In the first coherence interval, the pilot transmissions during phases A and B are correspondingly the same as those of the HD scheme.
In phase C, due to the ability of FD, the RS receives the source data as well as forwarding the downlink data to destinations.
Different from the HD overlay scheme, the downlink data forwarding during phase D is exactly overlapped by the source pilot transmission in phase A of the subsequent coherence interval.
The subsequent phases of the second coherence interval correspondingly repeats those of the first interval.
As for the third and following coherence intervals, the communication procedures are identical to those of the second interval.

\begin{remark}\label{rmk:Overhead}
	The pilot transmission overhead can be calculated by $\eta_p=T_p/T_c$, where $T^{\text{C}}_p=2K$ and $T^{\text{P}}_p=K$ for the conventional and proposed schemes, respectively.
	Hence, it is straightforward to obtain the overheads of the conventional and proposed schemes to be $\eta^{\text{C}}_p=2K/T_c$ and $\eta^{\text{P}}_p=K/T_c$, respectively.
	It is obvious that $\eta^{\text{C}}_p\gg\eta^{\text{P}}_p$, especially for the massive MIMO scenario where $K$ can be large.
\end{remark}

\begin{remark}\label{rmk:Throughput}
	For a non-buffered relaying system where the number of forwarding data exactly equals to that of the source data, the portion of data transmission within a coherence interval can be calculated as $\eta^{\text{HD}}_d=T^{\text{P}}_d/2T_c$ in the HD overlay system, where $T^{\text{P}}_d=T_c-K$.
	Compared to that, it is further increased to $\eta^{\text{FD}}_d=\mathcal{L}T^{\text{P}}_d/(\mathcal{L}T_\mathrm{c}+T^{\text{P}}_p)\approx(T_\mathrm{c}-K)/T_\mathrm{c}$ in the FD system,	where $\mathcal{L}$ is the total number of successive coherence intervals used for communications and the approximation is taken when $\mathcal{L}$ is large.
	FD almost doubles the efficiency of HD data transmission due to the FD property and the proposed pilot-data overlay structure.
	Note that the source pilot transmission is not indented to overlap the downlink data transmission of the previous coherence interval in the HD mode.
	It is argued that a TDD RS can not receive source pilots when the transceiver is working in the transmission mode.
	Therefore, the FD overlay scheme economizes more resources for data transmissions than the HD one.
\end{remark}

\section{Channel Estimation}\label{sec:ChannelEstimation}
In this section, the mathematical formulations of channel estimations with the proposed scheme are presented for both duplex relaying systems.

\subsection{Source Channel Estimation of the First Interval}\label{sec:SourceChannelEstimationFirstInterval}
In the proposed pilot-data overlay structure, the source channel estimations during the first coherence interval for both duplex systems are identical (see Figs. \ref{fig:PhaseDiagram}(c) and \ref{fig:PhaseDiagram}(d)).

In the first coherence interval, the RS receives source pilots without contamination during phase A while the transmitter of RS and all destination users keep mute.
Supposing source users send the pilot matrix $\bm{\Phi}\in\C^{K\times K}$ to RS with power $\rho_\mathrm{p}$ per user, where the $k$th row of the matrix, $\bm{\phi}_k$, is the pilot sequence sent by the $k$th source user and $\bm{\Phi}\bm{\Phi}^\mathrm{H}=\mathbf{I}_K$ due to orthogonality, the received signal at RS can be expressed as
\begin{equation}
	\mathbf{R}^\mathrm{A}[1] = \sqrt{K\rho_\mathrm{p}}\mathbf{G}_\mathrm{s}[1]\bm{\Phi} + \mathbf{N}^\mathrm{A}[1]
\end{equation}
where $\mathbf{N}^\mathrm{A}[1]$ is the additive white Gaussian noise (AWGN) matrix constructed by $\CN(0, 1)$ RVs.
By employing the MMSE criteria, the estimate of source channels can be obtained as
\begin{equation}\label{eq:SourceChannelEstimationFirstInterval}
	\mathbf{\hat{G}}_\mathrm{s}[1] = \frac{1}{\sqrt{K\rho_\mathrm{p}}}\mathbf{R}^\mathrm{A}[1]\bm{\Phi}^\Hm\mathbf{\tilde{D}}_\mathrm{s}[1]
\end{equation}
where $\mathbf{\tilde{D}}_\mathrm{s}[1]\triangleq(\mathbf{I}_K+\frac{1}{K\rho_\mathrm{p}}\mathbf{D}_\mathrm{s}^{-1})^{-1}$.
Due to the property of MMSE estimation, the channel matrix can be decomposed into two independent components as
\begin{equation}\label{eq:SourceChannelMatrixDecompositionFirstInterval}
	\mathbf{G}_\mathrm{s}[1] = \mathbf{\hat{G}}_\mathrm{s}[1] + \mathcal{E}_\mathrm{s}[1]
\end{equation}
where $\mathcal{E}_\mathrm{s}[1]$ is the error matrix constructed by columns mutually independent of the corresponding column entries of $\mathbf{\hat{G}}_\mathrm{s}[1]$, mathematically,
\begin{equation}\label{eq:SourceChannelVectorDecompositionFirstInterval}
	\mathbf{g}_{\mathrm{s}k}[1] = \mathbf{\hat{g}}_{\mathrm{s}k}[1] + \bm{\varepsilon}_{\mathrm{s}k}[1]
\end{equation}
where $\mathbf{\hat{g}}_{\mathrm{s}k}[1]$ and $\bm{\varepsilon}_{\mathrm{s}k}[1]$ are the $k$th ($k=1,2,\cdots,K$) column vector of $\mathbf{\hat{G}}_\mathrm{s}[1]$ and $\mathcal{E}_\mathrm{s}[1]$, respectively, $\mathbf{\hat{g}}_{\mathrm{s}k}[1]\sim\CN(\mathbf{0}, \sigma_{\mathrm{s}k}^2[1]\mathbf{I}_M)$, $\bm{\varepsilon}_{\mathrm{s}k}[1]\sim\CN(\mathbf{0}, \varepsilon_{\mathrm{s}k}^2[1]\mathbf{I}_M)$, $\sigma_{\mathrm{s}k}^2[1]= K\rho_\mathrm{p}\beta_{\mathrm{s}k}^2/(1+K\rho_\mathrm{p}\beta_{\mathrm{s}k})$, and $\varepsilon_{\mathrm{s}k}^2[1]\triangleq\beta_{\mathrm{s}k}-\sigma_{\mathrm{s}k}^2[1]$.

\subsection{Source Channel Estimation of Subsequent Intervals}\label{sec:SourceChannelEstimationSuccessiveIntervals}
According to Remark \ref{rmk:Throughput} and Fig. \ref{fig:PhaseDiagram}(c), the scenario of HD source pilot transmission (phase A) is identical for all coherence intervals and hence the HD source channel estimation is fully addressed in Subsection \ref{sec:SourceChannelEstimationFirstInterval}. 
However, the story differs regarding the FD where the transmitter of RS is working on forwarding the downlink data (phase D) while the receiver is receiving source pilots simultaneously (phase A) in the second coherence interval and after.
Therefore, the RS receives both the source pilots and downlink data leakages, which means that the source pilot is contaminated by LI.
This subsection considers the source channel estimation of the FD mode in the second and succeeding coherence intervals and characterizes the estimation errors introduced by both AWGN and LI.

Without loss of generality, take the $\iota$th ($\iota>1$) coherence time interval for instance.
The received source pilots at RS can be expressed as
\begin{equation}
	\begin{split}
		\mathbf{R}^\mathrm{A}[\iota] = &\underbrace{\sqrt{K\rho_\mathrm{p}}\mathbf{G}_\mathrm{s}[\iota]\bm{\Phi}}_{\text{desired signal}} + \\ 
								       &\underbrace{\sqrt{\rho_\mathrm{d}}\alpha[\iota-1]\mathbf{G}_\mathrm{LI}\mathbf{\hat{G}}_\mathrm{d}[\iota-1]\mathbf{X}^\mathrm{D}[\iota-1]}_{\text{LI}} + \underbrace{\mathbf{N}^\mathrm{A}[\iota]}_{\text{AWGN}}
	\end{split}
\end{equation}
where $\rho_\mathrm{p}$ and $\rho_\mathrm{d}$ represent the transmission power of source pilots and RS forwarding data, respectively,
$\mathbf{\hat{G}}_\mathrm{d}[\iota-1]$ (given by (\ref{eq:DestinationChannEst})) denotes the MRT precoding matrix of the forwarding data $\mathbf{X}^\mathrm{D}[\iota-1]$ with a power normalization factor $\alpha[\iota-1]$, and $\mathbf{N}^\mathrm{A}[\iota]\in\mathbb{C}^{M\times K}$ is the noise matrix constructed by $\CN(0, 1)$ RVs.
For detailed descriptions of the LI term and the expression of $\alpha[\iota-1]$ refer to Sections \ref{sec:DownlinkAnalysis} and \ref{sec:AnalysisOfUplinkPhaseC}.
By applying MMSE channel estimation, the estimate of the source channels is obtained by
\begin{equation}\label{eq:SourceChannEst}
	\mathbf{\hat{G}}_\mathrm{s}[\iota]=\frac{1}{\sqrt{K\rho_\mathrm{p}}}\mathbf{R}^\mathrm{A}[\iota]\bm{\Phi}^\mathrm{H}\mathbf{\tilde{D}}_\mathrm{s}[\iota]
\end{equation}
where $\mathbf{\tilde{D}}_\mathrm{s}[\iota]\triangleq\left(\mathbf{I}_K+\frac{\rho_\mathrm{d}\beta_\mathrm{LI}+1}{K\rho_\mathrm{p}}\mathbf{D}_\mathrm{s}^{-1}\right)^{-1}$.
Similarly, the channel matrix can also be decomposed into mutually independent two components as follows:
\begin{equation}
	\mathbf{G}_\mathrm{s}[\iota] = \mathbf{\hat{G}}_\mathrm{s}[\iota] + \mathcal{E}_\mathrm{s}[\iota]
\end{equation}
where $\mathcal{E}_\mathrm{s}[\iota]$ denotes the error matrix of estimations.
With regard to the $k$th columns of matrices $\mathbf{\hat{G}}_\mathrm{s}[\iota]$ and $\mathcal{E}_\mathrm{s}[\iota]$, there exist $\mathbf{\hat{g}}_{\mathrm{s}k}[\iota]\sim\mathcal{CN}\left(\mathbf{0}, \sigma_{\mathrm{s}k}^2[\iota]\mathbf{I}_M\right)$ and $\bm{\varepsilon}_{\mathrm{s}k}[\iota]\sim\mathcal{CN}\left(\mathbf{0}, \varepsilon_{\mathrm{s}k}^2[\iota]\mathbf{I}_M\right)$, where $\sigma_{\mathrm{s}k}^2[\iota]\triangleq{K\rho_\mathrm{p}\beta_{\mathrm{s}k}^2}/(\rho_\mathrm{d}\beta_\mathrm{LI}+1+K\rho_\mathrm{p}\beta_{\mathrm{s}k})$ and $\varepsilon_{\mathrm{s}k}^2[\iota]\triangleq\beta_{\mathrm{s}k}-\sigma_{\mathrm{s}k}^2[\iota]$, respectively, for $k$ from $1$ to $K$.

\subsection{Destination Channel Estimation}\label{sec:DestinationChannelEstimation}
In this subsection, the communication in phase B is formulated during which the source data and destination pilots are transmitted simultaneously.
To simplify the description, the source data is separated into two successive parts as $\mathbf{S}=[\mathbf{S}^\mathrm{B}~\mathbf{S}^\mathrm{C}]$, where $\mathbf{S}^\mathrm{B}\in\C^{K\times K}$ is the source data transmitted within phase B and $\mathbf{S}^\mathrm{C}\in\C^{K\times (T_\mathrm{d}-K)}$ is within phase C.
Here, $T_\mathrm{d}$ is the total length of the source data to transmit by each user within a coherence interval.
Without loss of generality, we assume $T_\mathrm{d}>K$.
The source data $\mathbf{S}^\mathrm{B}$ is transmitted alongside destination pilots transmission.
The following elaboration reveals that $\mathbf{S}^\mathrm{B}$ can be exactly detected from the received signal and the contamination to destination channel estimation can be suppressed by applying source data cancellation with a large number of RS antennas, due to the orthogonality between uplink and downlink channels.

At first, the RS detects the source data from the interfered received signal which can be expressed as
\begin{equation}\label{eq:ReceivedSignalSB}
	\mathbf{R}^\mathrm{B}[\iota] = \sqrt{\rho_\mathrm{s}}\mathbf{G}_\mathrm{s}[\iota]\mathbf{S}^\mathrm{B}[\iota] + \sqrt{K\rho_\mathrm{p}}\mathbf{G}_\mathrm{d}[\iota]\bm{\Psi} + \mathbf{N}^\mathrm{B}[\iota]
\end{equation}
where $\bm{\Psi}\in\C^{K\times K}$ denotes the destination pilot matrix transmitted at power $\rho_\mathrm{p}$ per user and $\mathbf{N}^\mathrm{B}[\iota]$ is the AWGN matrix consisting of $\CN(0, 1)$ RVs.
The MRC is applied to combine signals received by the RS antennas, where the combiner is $\mathbf{\hat{G}}_\mathrm{s}[\iota]$ given by (\ref{eq:SourceChannEst}).
Hence, the combined signal is
\begin{equation}\label{eq:DataDetectionSB}
	\mathbf{\tilde{S}}^\mathrm{B}[\iota] = \mathbf{\hat{G}}_\mathrm{s}^\Hm[\iota]\mathbf{R}^\mathrm{B}[\iota].
\end{equation}
Therefore, by employing the law of large numbers \cite{cramer2004random}, Proposition \ref{prp:SourceDataDetectionSB} summarizes the source data detection.

\begin{proposition}\label{prp:SourceDataDetectionSB}
	With MRC processing, the source data $\mathbf{S}^\mathrm{B}[\iota]$ can be exactly detected from $\mathbf{\tilde{S}}^\mathrm{B}[\iota]$, if a large number of reception antennas are equipped at RS, i.e., $M\rightarrow\infty$.
\end{proposition}
\begin{IEEEproof}
	See Appendix \ref{apdx:prf:prp:SourceDataDetectionSB}.
\end{IEEEproof}

Hereby, it is ready to estimate the destination channel by canceling the detected source data from the received signal to reduce pilot contaminations.
By recalling the received signal from (\ref{eq:ReceivedSignalSB}) and subtracting the product of the detected source signal and the hermitian of the estimated source channels, we obtain the MMSE estimation of the destination channels as
\begin{equation}\label{eq:DestinationChannEst}
	\mathbf{\hat{G}}_{\mathrm{d}}[\iota] = \frac{1}{\sqrt{K\rho_\mathrm{p}}}\left(\mathbf{R}^\mathrm{B}[\iota]-\sqrt{\rho_\mathrm{s}}\mathbf{\hat{G}}_\mathrm{s}[\iota]\mathbf{S}^\mathrm{B}[\iota]\right)\bm{\Psi}^\Hm\mathbf{\tilde{D}}_\mathrm{d}[\iota]
\end{equation}
where $\mathbf{\tilde{D}}_\mathrm{d}[\iota]\triangleq\left(\mathbf{I}_K+\frac{\rho_\mathrm{s}\sum_{i=1}^{K}\varepsilon_{\mathrm{s}i}^2[\iota]+1}{K\rho_\mathrm{p}}\mathbf{D}_\mathrm{d}^{-1}\right)^{-1}$.
And the MMSE estimation follows
\begin{equation}
	\mathbf{G}_{\mathrm{d}}[\iota] = \mathbf{\hat{G}}_{\mathrm{d}}[\iota] + \mathcal{E}_{\mathrm{d}}[\iota]
\end{equation}
where $\mathbf{\hat{G}}_{\mathrm{d}}[\iota]$ and $\mathcal{E}_{\mathrm{d}}[\iota]$ are independent of each other.
Particularly, the $k$th columns of both matrices, $\mathbf{\hat{g}}_{\mathrm{d}k}[\iota]$ and $\bm{\varepsilon}_{\mathrm{d}k}[\iota]$, are mutually independent random vectors, following distribution $\CN(\mathbf{0}, \sigma_{\mathrm{d}k}^2[\iota]\mathbf{I}_M)$ and $\CN(\mathbf{0}, \varepsilon_{\mathrm{d}k}^2[\iota]\mathbf{I}_M)$, respectively, where $\sigma_{\mathrm{d}k}^2[\iota]\triangleq K\rho_\mathrm{p}\beta_{\mathrm{d}k}^2/\left(\rho_\mathrm{s}\sum_{i=1}^{K}\varepsilon_{\mathrm{s}i}^2[\iota]+1 + K\rho_\mathrm{p}\beta_{\mathrm{d}k}\right)$ and $\varepsilon_{\mathrm{d}k}^2[\iota]\triangleq\beta_{\mathrm{d}k}-\sigma_{\mathrm{d}k}^2[\iota]$, for $k$ from $1$ to $K$.

\begin{remark}
	The covariance factor of the source channel estimate, $\sigma_{\mathrm{s}k}^2[\iota]$, is independent of the coherence interval index $\iota$.
	In addition, the factor of destination channel estimate, $\sigma_{\mathrm{d}k}^2[\iota]$, only depends on the source channel estimation errors, which are independent of $\iota$.
	From this phenomenon, it is interesting to note that no error propagation exists for the proposed pilot-data transmission scheme in both duplex relaying systems, which differs from the semi-orthogonal pilot design proposed in \cite{Zhang2014} where CSI estimation errors accumulate as the increase of $\iota$.
\end{remark}

\section{Achievable Rate Analysis}\label{sec:AchievableRateAnalysis}
This section characterizes the performance of the proposed pilot-data transmission scheme by evaluating achievable rates of the considered massive MIMO relaying systems.
For the multipair communication, the normalized system achievable rate is defined as an average of sum rates among all user pairs over the entire transmission time, that is
\begin{equation}\label{eq:OverallAchievableRate}
	\mathcal{R} = \frac{1}{\mathcal{L}T_\mathrm{c}}\sum_{\iota=1}^{\mathcal{L}}\sum_{k=1}^{K}\mathcal{R}_k[\iota].
\end{equation}
The individual achievable rate in the decode-and-forward (DF) relaying system is given by
\begin{equation}\label{eq:UserPairAchievableRate}
	\mathcal{R}_k[\iota] = \min\{\mathcal{R}^\mathrm{UL}_k[\iota], \mathcal{R}^\mathrm{DL}_k[\iota]\},
\end{equation}
where $\mathcal{R}^\mathrm{UL}_k[\iota]$ and $\mathcal{R}^\mathrm{DL}_k[\iota]$ denote the uplink and downlink rates between user pair $k$ and RS in the coherence interval $\iota$, respectively.
Here we employ the technique developed by \cite{Jose2011} to approximate the ergodic achievable rate for per-link communication, i.e., $\mathcal{R}^\mathrm{UL}_k[\iota]$ and $\mathcal{R}^\mathrm{DL}_k[\iota]$.
In this technique, the received signal is separated into desired signal and effective noise terms, where the former term is the product of transmitted signal and the expectation of channels while the latter one consists of uncorrelated interferences and AWGN.
Hence, only the statistical, other than instantaneous, CSI is required to evaluate the achievable rate.
The rate calculated by this technique is the lower bound of the exact one, and numerical results presented in both \cite{Hoydis2013} and \cite{Ngo2014} show that it is tolerably close to the genie result produced by Monte-Carlo simulation.
Consequently, the per-link ergodic achievable rate within a coherence interval is bounded by
\begin{equation}\label{eq:e2eAchievableRate}
	\mathcal{R}_k^{\mathrm{PL}}[\iota] = \tau_\mathrm{d}[\iota]\log_2(1+\gamma_k^{\mathrm{PL}}[\iota])
\end{equation}
where $\tau_\mathrm{d}[\iota]$ denotes the data transmission time within the coherence interval $\iota$ and the effective signal to noise ratio is defined as $\gamma_k^{\mathrm{PL}}[\iota]\triangleq\mathcal{S}_k^{\mathrm{PL}}[\iota] / (\mathcal{I}_k^{\mathrm{PL}}[\iota] + \mathcal{N}_k^{\mathrm{PL}}[\iota])$.
Here, $\mathcal{S}_k^{\mathrm{PL}}[\iota]$, $\mathcal{I}_k^{\mathrm{PL}}[\iota]$ and $\mathcal{N}_k^{\mathrm{PL}}[\iota]$ represent the power of the desired signal, uncorrelated interference and AWGN, respectively.

\subsection{Downlink Analysis}\label{sec:DownlinkAnalysis}
Here the downlink achievable rates for both HD and FD schemes are analyzed.
By applying the MRT processing at the RS to the downlink data $\mathbf{X}[\iota]\in\C^{K\times T_\mathrm{d}}$ and transmitting to the destination channels with power $\rho_\mathrm{d}$, the received signal at destination users is obtained, for user $k$ ($k=1,2,\cdots,K$), as
\begin{equation}\label{eq:DownlinkReceivedSignal}
	\mathbf{y}_k[\iota] = \sqrt{\rho_\mathrm{d}}\alpha[\iota]\mathbf{g}_{\mathrm{d}k}^\Hm[\iota]\mathbf{\hat{G}}_\mathrm{d}[\iota]\mathbf{X}[\iota] + \mathbf{z}_k[\iota]
\end{equation}
where $\mathbf{z}_k[\iota]\in\C^{1\times T_\mathrm{d}}$ is the AWGN vector consisting of $\CN(0,1)$ RVs and $\alpha[\iota]$ is the factor to normalize the average transmit power, i.e., letting $\E\{\|\alpha[\iota]\mathbf{\hat{G}}_\mathrm{d}[\iota]\|^2\}=1$, thus $\alpha[\iota]=\sqrt{1\big/\left(M\sum_{i=1}^{K}\sigma_{\mathrm{d}i}^2[\iota]\right)}$.
To separate the desired signal from the interference and noise, (\ref{eq:DownlinkReceivedSignal}) can be rewritten as
\begin{equation}\label{eq:ReceivedSignalDecomposeX}
	\mathbf{y}_k[\iota]=\underbrace{\sqrt{\rho_\mathrm{d}}\alpha[\iota]\E\left\{\mathbf{g}_{\mathrm{d}k}^\Hm[\iota]\mathbf{\hat{g}}_{\mathrm{d}k}[\iota]\right\}\mathbf{x}_k[\iota]}_{\text{desired signal}} + \underbrace{\mathbf{\breve{z}}_k[\iota]}_{\text{effective noise}}
\end{equation}
where the effective noise is defined by
\[
	\begin{split}
		\mathbf{\breve{z}}_k[\iota] \triangleq& \sqrt{\rho_\mathrm{d}}\alpha[\iota]\left(\mathbf{g}_{\mathrm{d}k}^\Hm[\iota]\mathbf{\hat{g}}_{\mathrm{d}k}[\iota]-\E\left\{\mathbf{g}_{\mathrm{d}k}^\Hm[\iota]\mathbf{\hat{g}}_{\mathrm{d}k}[\iota]\right\}\right)\mathbf{x}_k[\iota]\\
		& + \sqrt{\rho_\mathrm{d}}\alpha[\iota]\sum_{i=1,i\neq k}^{K}\mathbf{g}_{\mathrm{d}k}^\Hm[\iota]\mathbf{\hat{g}}_{\mathrm{d}i}[\iota]\mathbf{x}_i[\iota] + \mathbf{z}_k[\iota].
	\end{split}
\]
Therefore, the effective SINR of the received signal at the $k$th destination user can be expressed as
\begin{equation}\label{eq:GammaDownlinkOrigin}
	\gamma_k^{\mathrm{DL}}[\iota] = \frac{\rho_\mathrm{d}\alpha^2[\iota]\left|\E\left\{\mathbf{g}_{\mathrm{d}k}^\Hm[\iota]\mathbf{\hat{g}}_{\mathrm{d}k}[\iota]\right\}\right|^2}{\rho_\mathrm{d}\alpha^2[\iota]\Var\left\{\mathbf{g}_{\mathrm{d}k}^\Hm[\iota]\mathbf{\hat{g}}_{\mathrm{d}k}[\iota]\right\} + \mathrm{MI}_k^{\mathrm{DL}}[\iota] + 1}
\end{equation}
where the power of the downlink multipair interference (MI) is defined by
\begin{equation}\label{eq:MIDownlink}
	\mathrm{MI}_k^{\mathrm{DL}}[\iota]\triangleq\rho_\mathrm{d}\alpha^2[\iota]\sum_{i=1,i\neq k}^{K}\E\left\{\left|\mathbf{g}_{\mathrm{d}k}^\Hm[\iota]\mathbf{\hat{g}}_{\mathrm{d}i}[\iota]\right|^2\right\}.
\end{equation}

\begin{theorem}\label{thm:DownlinkAchievableRate}
	By employing the MRT processing, the achievable rate of the downlink data forwarded to the destination user $k$ ($k=1,2,\cdots,K$) in both HD and FD, for a finite number of RS transmitter antennas $M$, can be characterized by
	\begin{equation}\label{eq:DownlinkAchievableRate}
		\mathcal{R}^\mathrm{DL}_k[\iota] = T_d\log_2\left(1+\gamma_k^{\mathrm{DL}}[\iota]\right)
	\end{equation}
	where
	\begin{equation}\label{eq:GammaDownlink}
		\gamma_k^\mathrm{DL}[\iota] = \frac{M\sigma_{\mathrm{d}k}^4[\iota]}{(\beta_{\mathrm{d}k}+1/\rho_\mathrm{d})\sum_{i=1}^{K}\sigma_{\mathrm{d}i}^2[\iota]}.
	\end{equation}
\end{theorem}
\begin{IEEEproof}
	See Appendix \ref{apdx:prf:thm:DownlinkAchievableRate}.
\end{IEEEproof}

\subsection{Uplink Analysis}
In the proposed pilot-data overlay scheme, the uplink data is transmitted in two successive phases where the first part of data is transmitted during phase B and the remaining is sent within phase C.
For the two duplex systems, the phase B communication is similar while the phase C differs.
The following description first conducts the rate analysis of phase B for both duplex systems, and then perform the phase C analysis distinguished by each mode.

\subsubsection{Analysis of uplink phase B}
The $k$th row of $\mathbf{\tilde{S}}^\mathrm{B}[\iota]$ in (\ref{eq:DataDetectionSB}) can be rewritten as
\begin{equation}\label{eq:ReceivedSignalDecomposeSB}
	\mathbf{\tilde{s}}_k^\mathrm{B}[\iota]=\underbrace{\sqrt{\rho_\mathrm{s}}\E\left\{\mathbf{\hat{g}}_{\mathrm{s}k}^\Hm[\iota]\mathbf{g}_{\mathrm{s}k}[\iota]\right\}\mathbf{s}^\mathrm{B}_k[\iota]}_{\text{desired signal}}+\underbrace{\mathbf{\breve{n}}^\mathrm{B}_k[\iota]}_{\text{effective noise}}
\end{equation}
where the effective noise is
\[
	\begin{split}
		\mathbf{\breve{n}}^\mathrm{B}_k[\iota]\triangleq& \sqrt{\rho_\mathrm{s}}\left(\mathbf{\hat{g}}_{\mathrm{s}k}^\Hm[\iota]\mathbf{g}_{\mathrm{s}k}[\iota]-\E\left\{\mathbf{\hat{g}}_{\mathrm{s}k}^\Hm[\iota]\mathbf{g}_{\mathrm{s}k}[\iota]\right\}\right)\mathbf{s}^\mathrm{B}_k[\iota]\\
		&+ \sqrt{\rho_\mathrm{s}}\sum_{i=1,i\neq k}^{K}\mathbf{\hat{g}}_{\mathrm{s}k}^\Hm[\iota]\mathbf{g}_{\mathrm{s}i}[\iota]\mathbf{s}^\mathrm{B}_i[\iota]\\
		&+ \sqrt{K\rho_\mathrm{p}}\mathbf{\hat{g}}_{\mathrm{s}k}^\Hm[\iota]\mathbf{G}_{\mathrm{d}}[\iota]\bm{\Psi} 
		+ \mathbf{\hat{g}}_{\mathrm{s}k}^\Hm[\iota]\mathbf{N}^\mathrm{B}[\iota].
	\end{split}
\]
Then, the effective SINR of the received signal during phase B can be expressed by
\begin{equation}\label{eq:GammaUplinkOriginB}
	\gamma_{k}^\mathrm{B}[\iota]=\frac{\rho_\mathrm{s}\left|\E\left\{\mathbf{\hat{g}}_{\mathrm{s}k}^\Hm[\iota]\mathbf{g}_{\mathrm{s}k}[\iota]\right\}\right|^2}{\rho_\mathrm{s}\Var\left\{\mathbf{\hat{g}}_{\mathrm{s}k}^\Hm[\iota]\mathbf{g}_{\mathrm{s}k}[\iota]\right\}+\mathrm{MI}_k^\mathrm{UL}[\iota]+\mathrm{PI}_k^\mathrm{UL}[\iota]+\mathrm{AN}_k^\mathrm{UL}[\iota]}
\end{equation}
where the power of uplink MI, destination pilot interference (PI) and AWGN are respectively defined as
\begin{align}
	\label{eq:UplinkMI}
	\mathrm{MI}_k^\mathrm{UL}[\iota]&\triangleq\rho_\mathrm{s}\sum_{i=1,i\neq k}^K\E\left\{\left|\mathbf{\hat{g}}_{\mathrm{s}k}^\Hm[\iota]\mathbf{g}_{\mathrm{s}i}[\iota]\right|^2\right\} \\
	\label{eq:UplinkPI}
	\mathrm{PI}_k^\mathrm{UL}[\iota]&\triangleq\rho_\mathrm{p}\E\left\{\left\|\mathbf{\hat{g}}_{\mathrm{s}k}^\Hm[\iota]\mathbf{G}_{\mathrm{d}}[\iota]\right\|^2\right\}\\
	\label{eq:UplinkAN}
	\mathrm{AN}_k^\mathrm{UL}[\iota]&\triangleq\E\left\{\left\|\mathbf{\hat{g}}_{\mathrm{s}k}[\iota]\right\|^2\right\}.
\end{align}

\subsubsection{Analysis of uplink phase C}\label{sec:AnalysisOfUplinkPhaseC}
Regarding the uplink data transmission during phase C, only MI interferes for the HD communication while for the FD, LI also exists.
The received signal at RS during phase C can be expressed as
\begin{equation}\label{eq:ReceivedSignalSC}
	\mathbf{R}^\mathrm{C}[\iota]=\sqrt{\rho_\mathrm{s}}\mathbf{G}_\mathrm{s}[\iota]\mathbf{S}^\mathrm{C}[\iota]+\mathbf{L}^\mathrm{C}[\iota]+\mathbf{N}^\mathrm{C}[\iota]
\end{equation}
where $\mathbf{L}^\mathrm{C}[\iota]\triangleq\sqrt{\rho_\mathrm{d}}\alpha[\iota]\mathbf{G}_\mathrm{LI}[\iota]\mathbf{\hat{G}}_\mathrm{d}^\Hm[\iota]\mathbf{X}^\mathrm{C}[\iota]$ is the LI for the FD relaying while for HD, it is zero, and $\mathbf{N}^\mathrm{C}[\iota]$ is the AWGN matrix.
By applying MRC to the $k$th source data with similar manipulations to the combined signal as in (\ref{eq:ReceivedSignalDecomposeSB}), the effective SINR of the received signal during phase C is obtained as
\begin{equation}\label{eq:GammaUplinkOriginC}
	\gamma_{k}^\mathrm{C}[\iota]=\frac{\rho_\mathrm{s}\left|\E\left\{\mathbf{\hat{g}}_{\mathrm{s}k}^\Hm[\iota]\mathbf{g}_{\mathrm{s}k}[\iota]\right\}\right|^2}{\rho_\mathrm{s}\Var\left\{\mathbf{\hat{g}}_{\mathrm{s}k}^\Hm[\iota]\mathbf{g}_{\mathrm{s}k}[\iota]\right\}+\mathrm{MI}_k^\mathrm{UL}[\iota]+\mathrm{LI}_k^\mathrm{UL}[\iota]+\mathrm{AN}_k^\mathrm{UL}[\iota]}
\end{equation}
where $\mathrm{MI}_k^\mathrm{UL}[\iota]$ and $\mathrm{AN}_k^\mathrm{UL}[\iota]$ are respectively defined by (\ref{eq:UplinkMI}) and (\ref{eq:UplinkAN}), and the power of LI is defined as
\begin{equation}\label{eq:UplinkLI}
	\mathrm{LI}_k^\mathrm{UL}[\iota]\triangleq\begin{cases}
	0, & \text{HD}\\
	\rho_\text{d}\alpha^2[\iota]\E\left\{\left\|\mathbf{\hat{g}}_{\text{s}k}^\text{H}[\iota]\mathbf{G}_{\text{LI}}[\iota]\mathbf{\hat{G}}_\mathrm{d}[\iota]\right\|^2\right\}, & \text{FD}.
	\end{cases}
\end{equation}

Finally, the uplink achievable rates of pilot-data overlay relaying systems are given by Theorem \ref{thm:UplinkAchievableRate}.
\begin{theorem}\label{thm:UplinkAchievableRate}
	With MRC processing, the uplink achievable rate of source user $k$ ($k=1,2,\cdots K$) during coherence interval $\iota$ in both HD and FD systems, for a finite antenna number $M$, can be characterized by
	\begin{equation}\label{eq:UplinkAchievableRate}
		\mathcal{R}_k^\mathrm{UL}[\iota]=K\log_2(1+\gamma_k^\mathrm{B}[\iota])+(T_\mathrm{d}-K)\log_2(1+\gamma_k^\mathrm{C}[\iota])
	\end{equation}
	where 
	\begin{equation}\label{eq:GammaUplinkB}
		\gamma_k^\mathrm{B}[\iota] = \frac{M\sigma_{\mathrm{s}k}^2[\iota]}{\sum_{i=1}^{K}\beta_{\mathrm{s}i}+\left(\rho_\mathrm{p}\sum_{i=1}^{K}\beta_{\mathrm{d}i}+1\right)/\rho_\mathrm{s}}
	\end{equation}
	and
	\begin{equation}\label{eq:GammaUplinkC}
		\gamma_k^\mathrm{C}[\iota] = \begin{cases}
			\frac{M\sigma_{\mathrm{s}k}^2[\iota]}{\sum_{i=1}^{K}\beta_{\mathrm{s}i} + 1/\rho_\mathrm{s}}, & \text{HD}\\
			\frac{M\sigma_{\mathrm{s}k}^2[\iota]}{\sum_{i=1}^{K}\beta_{\mathrm{s}i} + \left(\rho_\mathrm{d}\beta_\mathrm{LI}+1\right)/\rho_\mathrm{s}}, & \text{FD}.
		\end{cases}
	\end{equation}
\end{theorem}
\begin{IEEEproof}
	See Appendix \ref{apdx:prf:thm:UplinkAchievableRate}
\end{IEEEproof}

\begin{remark}
	Having the numerator and denominator of the FD case in (\ref{eq:GammaUplinkC}) both divided by $M$, it is observed that the LI term, $\rho_\mathrm{d}\beta_\mathrm{LI}/M$, vanishes as the number of RS antennas tending infinity, i.e., $M\to\infty$.
	This is consistent with the observation on LI discovered in \cite{Ngo2014a} but for a different pilot-data transmission scheme.
\end{remark}

\section{Asymptotic Analysis}\label{sec:AsymptoticAnalysis}
This section analyzes the asymptotic performance of the proposed scheme at high and low SNR regions in comparison to the conventional pilot scheme under the assumption that a large number of user pairs are served by large-scale RS antennas.
In following asymptotic derivations, it is assumed that the transmit power of all data and pilot is identical, i.e., $\rho_\text{s}=\rho_\text{d}=\rho_\text{p}=\rho$.

Firstly, the asymptotic performance at the high SNR region is analyzed with the proposed pilot scheme.
By setting $\rho\to\infty$, the effective SINRs denoted by (\ref{eq:GammaUplinkB}), (\ref{eq:GammaUplinkC}) and (\ref{eq:GammaDownlink}) can be written as
\begin{subequations}
	\begin{align}
		\label{eq:AsympHighSNRGammaUplinkB}
		\lim_{\rho\to\infty}\gamma_k^\mathrm{B}[\iota]=&\frac{M\beta_{\text{s}k}}{\sum_{i=1}^{K}\beta_{\text{s}i}+\sum_{i=1}^{K}\beta_{\text{d}i}}\\
		\label{eq:AsympHighSNRGammaUplinkC}
		\lim_{\rho\to\infty}\gamma_k^\mathrm{C}[\iota]=&\frac{M\beta_{\text{s}k}}{\sum_{i=1}^{K}\beta_{\text{s}i}}
	\end{align}
\end{subequations}
and
\begin{equation}\label{eq:AsympHighSNRGammaDownlink}
	\lim_{\rho\to\infty}\gamma_k^\mathrm{DL}[\iota]=\frac{M\beta_{\text{d}k}}{\sum_{i=1}^{K}\beta_{\text{d}i}}.
\end{equation}
In the above derivations, the channel estimate covariance factors are calculated as $\lim\limits_{\rho\to\infty}\sigma_{\mathrm{s}k}^2[\iota]=\beta_{\mathrm{s}k}$ and $\lim\limits_{\rho\to\infty}\sigma_{\text{d}k}^2[\iota]=\beta_{\text{d}k}$.

On the other hand, the low SNR region is considered by assuming $\rho\to 0$.
It is straightforward to obtain the effective SINRs of the uplink data transmission at low SNR as
\begin{equation}\label{eq:SPLowSNRUplinkSINR}
	\lim_{\rho\to 0}\gamma_k^\mathrm{B}[\iota] = \lim_{\rho\to 0}\gamma_k^\mathrm{C}[\iota] = MK\rho^2\beta_{\text{s}k}^2
\end{equation}
where $\lim\limits_{\rho\to 0}\sigma_{\text{s}k}^2[\iota]=K\rho\beta_{\text{s}k}^2$.
With similar manipulations, the downlink effective SINR expressed by (\ref{eq:GammaDownlink}) can be reformulated as
\begin{equation}\label{eq:hd:SPLowSNRDownlinkSINR}
	\lim_{\rho\to 0}\gamma_k^\mathrm{DL}[\iota] = \frac{MK\rho^2\beta_{\text{d}k}^4}{\sum_{i=1}^{K}\beta_{\text{d}i}^2}
\end{equation}
where $\lim\limits_{\rho\to 0}\varepsilon_{\text{s}k}^2[\iota]=\beta_{\text{s}k}-K\rho\beta_{\text{s}k}^2$ and $\lim\limits_{\rho\to 0}\sigma_{\text{d}k}^2[\iota]=K\rho\beta_{\text{d}k}^2$.

By respectively substituting the above asymptotic effective SINRs into (\ref{eq:UplinkAchievableRate}) and (\ref{eq:DownlinkAchievableRate}), the asymptotic per-link achievable rates are obtained, and hence the asymptotic overall system rate by summing them up.

For comparisons, the asymptotic achievable rates with the conventional pilot scheme at both high and low SNRs are presented as follows:
\begin{subequations}
	\begin{align}
		\label{eq:OPHighSNRUplinkSINR}
		\lim_{\rho\to\infty}\tilde{\gamma}_k^\mathrm{UL} =&\frac{M\beta_{\text{s}k}}{\sum_{i=1}^{K}\beta_{\text{s}i}} \\
		\label{eq:OPHighSNRDownlinkSINR}
		\lim_{\rho\to\infty}\tilde{\gamma}_k^\mathrm{DL} =& \frac{M\beta_{\text{d}k}}{\sum_{i=1}^{K}\beta_{\text{d}i}} \\
		\label{eq:OPLowSNRUplinkSINR}
		\lim_{\rho\to 0}\tilde{\gamma}_k^\mathrm{UL} =& MK\rho^2\beta_{\text{s}k}^2 \\
		\label{eq:OPLowSNRDownlinkSINR}
		\lim_{\rho\to 0}\tilde{\gamma}_k^\mathrm{DL} =& \frac{MK\rho^2\beta_{\text{d}k}^4}{\sum_{i=1}^{K}\beta_{\text{d}i}^2}.
	\end{align}
\end{subequations}
Note that $\tilde{(\cdot)}$ is used to indicate the conventional scheme in this section.

By comparing the achievable rates of two pilot schemes via the effective SINRs computed above, Corollary \ref{cry:AsymptoticResults} is drawn to confirm the advantages of the proposed scheme.

\begin{corollary}\label{cry:AsymptoticResults}
	With a large number of user pairs served by RS, the proposed pilot-data overlay transmission scheme outperforms the conventional one in multipair HD and FD massive MIMO relaying systems for both high and low SNRs.
\end{corollary}
\begin{IEEEproof}
	See Appendix \ref{apdx:prf:cry:AsymptoticResults}.
\end{IEEEproof}

\begin{remark}\label{rmk:IndepentOfCoInterval}
	According to Corollary \ref{cry:AsymptoticResults}, it is interesting to stress that the conclusion is independent of the coherence interval length $T_\mathrm{c}$ and the number of successive coherence intervals used for continuous communications $\mathcal{L}$.
	Therefore, the proposed scheme is always superior to, at least not worse than, the conventional scheme for both low and high SNRs, regardless of the coherence interval length.
\end{remark}

\section{Power Allocation for FD Relaying}\label{sec:PowerAllocation}
By analyzing the effective SINRs of the FD relaying system, one can find that the self-interference between uplink and downlink data transmissions, and hence the impact upon pilots sending, is a major factor that limits the system achievable rate.
Therefore, making a tradeoff between uplink and downlink data transmission is necessary to improve the overall system performance.
The power allocation is performed in the FD relaying system to optimize the transmit power of both source and forwarding data to control interference among uplink, downlink and pilot transmissions.
Note that this paper only considers balancing the source and RS data transmission power level while the power allocation for each individual source user is not taken into account.
In addition, this paper assume the total power consumption of all data transmission to be constrained under a fixed pilot transmit power.
In other words, data transmission power $\rho_\text{s}$ and $\rho_\text{d}$ are balanced with respect to a given $\rho_\text{p}$ to achieve the maximal overall system rate.

To begin with, the power allocation problem is formulated as
\begin{equation}\label{opt:fd:PowAlloc1}
	\begin{aligned}
		& \underset{\rho_\text{s},\rho_\text{d}}{\text{maximize}}& & \sum_{\iota=1}^{\mathcal{L}}\sum_{k=1}^{K}\mathcal{R}_k[\iota] \\
		& \text{subject to:}& & \mathcal{R}_k[\iota] = \min\left\{\mathcal{R}_k^{\text{UL}}[\iota], \mathcal{R}_k^{\text{DL}}[\iota]\right\}\\
		& & & \mathcal{L}T_d\left(K\rho_\text{s} + \rho_\text{d}\right) = E_\mathrm{d}\\
		& & & k=1,2,\cdots,K \text{ and }\iota=1,2,\cdots,\mathcal{L}
	\end{aligned}
\end{equation}
where $\mathcal{R}_k^{\text{UL}}[\iota]$, $\mathcal{R}_k^{\text{DL}}[\iota]$ and $\mathcal{R}_k[\iota]$ denote the uplink, downlink and end-to-end (e2e) achievable rate between the $k$th user pair and RS within the $\iota$th coherence time interval, respectively, and $E_\mathrm{d}$ is the total data transmission energy.
The operator $\min\{\cdot,\cdot\}$ in the above formulation is due to the DF relaying expressed in (\ref{eq:UserPairAchievableRate}), where the e2e achievable rate is determined by the minimum of the corresponding uplink and downlink rates.
To remove the minimum operator, the first constraint of problem (\ref{opt:fd:PowAlloc1}) can be reformulated equivalently to
\begin{equation}\label{opt:fd:ConstraintsIneq1}
	\begin{aligned}
		\mathcal{R}_k[\iota] &\leq \mathcal{R}_k^{\text{UL}}[\iota]\\
		\mathcal{R}_k[\iota] &\leq \mathcal{R}_k^{\text{DL}}[\iota].
	\end{aligned}
\end{equation}
By examining $\mathcal{R}_k^{\text{UL}}[\iota]$ and $\mathcal{R}_k^{\text{DL}}[\iota]$ with respect to $\rho_\text{s}$ and $\rho_\text{d}$, it is clear that these two inequality constraints are non-convex and thus it is difficult to solve (\ref{opt:fd:PowAlloc1}) directly.

To cope with this problem, a successive convex approximation (SCA) approach \cite{Moon2000,Hasna2004,Chiang2007} is proposed in this paper by replacing constraint (\ref{opt:fd:ConstraintsIneq1}) with linear inequalities.
That is to approximate both uplink and downlink rates by their first-order Taylor series in the $i$th iteration with the known power allocation $\bm{\rho}^{(i)}$ via
\begin{equation}\label{opt:fd:ConstraintsIneq2}
	\mathcal{R}^{\text{UL,DL}}_k[\iota](\bm{\rho})=\mathcal{R}^{\text{UL,DL}}_k[\iota](\bm{\rho}^{(i)})+\nabla^\text{T}\mathcal{R}^{\text{UL,DL}}_k[\iota](\bm{\rho}^{(i)})(\bm{\rho}-\bm{\rho}^{(i)})
\end{equation}
where $\bm{\rho}\triangleq[\rho_\text{s}~\rho_\text{d}]^\text{T}$ and the superscript $(\cdot)^{(i)}$ denotes the corresponding value at the $i$th iteration.
Substituting (\ref{opt:fd:ConstraintsIneq2}) into (\ref{opt:fd:ConstraintsIneq1}), the optimization problem can be reformulated in the $i$th iteration as
\begin{equation}\label{opt:fd:PowAlloc2}
	\begin{aligned}
		& \underset{\rho_\text{s},\rho_\text{d}}{\text{maximize}}& & \sum_{\iota=1}^{\mathcal{L}}\sum_{k=1}^{K}\mathcal{R}^{(i)}_k[\iota] \\
		& \text{subject to:}& & \mathcal{R}^{(i)}_k[\iota] \leq \mathcal{R}^{\text{UL}}_k[\iota](\bm{\rho}^{(i)})+\nabla^\text{T}\mathcal{R}^{\text{UL}}_k[\iota](\bm{\rho}^{(i)})(\bm{\rho}-\bm{\rho}^{(i)})\\
		& & & \mathcal{R}^{(i)}_k[\iota] \leq \mathcal{R}^{\text{DL}}_k[\iota](\bm{\rho}^{(i)})+\nabla^\text{T}\mathcal{R}^{\text{DL}}_k[\iota](\bm{\rho}^{(i)})(\bm{\rho}-\bm{\rho}^{(i)})\\
		& & & \mathcal{L}T_d\left(K\rho_\text{s} + \rho_\text{d}\right) = E_\mathrm{d}\\
		& & & k=1,2,\cdots,K \text{ and }\iota=1,2,\cdots,\mathcal{L}.
	\end{aligned}
\end{equation}
It is obvious that (\ref{opt:fd:PowAlloc2}) is a linear programming (LP) which can be solved efficiently by utilizing, e.g., the conventional interior-point method \cite{Boyd2004}.

The LP in (\ref{opt:fd:PowAlloc2}) is solved repeatedly in iterations with $i$ increased by 1 each time, until the increment of $\bm{\rho}^{(i)}$ is smaller than a set error tolerance.
After all, the $\bm{\rho}^{(i)}$ in the last iteration is the optimal power allocation $\bm{\rho}^*$ that leads to the maximum sum rate.
The SCA approach is summarized by Algorithm \ref{algtm:fd:SCA}.

The complexity and convergence analyses of the proposed SCA approach will be discussed in the numerical results of the next section.

\begin{algorithm} 
	\caption{Successive Convex Approximation Approach}
	\label{algtm:fd:SCA}
	\begin{algorithmic}[1]
		\REQUIRE Solve optimization problem (\ref{opt:fd:PowAlloc1})
		\INPUT $K$, $M$, $\mathcal{L}$, $\rho_\mathrm{p}$, $\rho_{\mathrm{LI}}$, $\beta_{\mathrm{s}k}$ and $\beta_{\mathrm{d}k}$
		\OUTPUT Optimized power allocation $\rho_\mathrm{s}^*$ and $\rho_\mathrm{d}^*$
		\STATE Initialize $\epsilon$ and set $i=0$ 
		\STATE Initialize $\bm{\rho}^{(0)}$
		\LOOP
		\STATE Solve the LP in (\ref{opt:fd:PowAlloc2}) and obtain the optimizer $\bm{\rho}^*$
		\IF{$\frac{\|\bm{\rho}^{*}-\bm{\rho}^{(i)}\|}{\|\bm{\rho}^{(i)}\|}<\epsilon$}
		\STATE Stop \textbf{loop}
		\ELSE
		\STATE $\bm{\rho}^{(i+1)}=\bm{\rho}^*$
		\STATE $i=i+1$
		\ENDIF
		\ENDLOOP
		\STATE Output $\rho_\mathrm{s}^*$ and $\rho_\mathrm{d}^*$.
	\end{algorithmic}
\end{algorithm}

\section{Numerical and Simulation Results}\label{sec:NumericalResults}
In this section, the performance of the proposed pilot-data overlay transmission scheme and the power allocation approach are studied by simulation and numerical evaluation for both HD and FD relaying systems.
Unless otherwise specified, by default, the RS is equipped with 128 antennas serving 10 pairs of users, $\rho_\mathrm{LI}$ is set to be 3 dB over the noise floor, the processing delay of the FD relaying for the conventional pilot scheme is 1 symbol slot, the coherence time interval is set to be 40 time slots, the transmission power of pilots, source data and forwarding data satisfy $\rho_\text{p} = \rho_\text{s} = \rho_\text{d}=20$ dB, and the Monte Carlo results are obtained by taking the average of 1000 random simulations with instantaneous channels.

\begin{figure}[htbp]
	\centering
	\includegraphics[width=\linewidth]{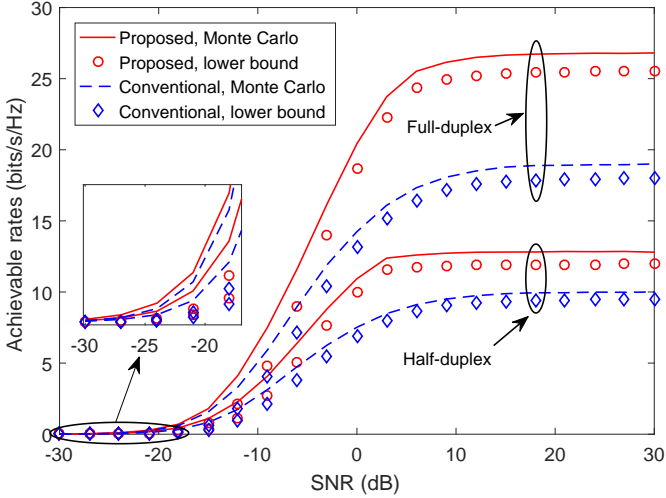}
	\caption{Comparisons of achievable rates between the proposed and conventional schemes under different SNRs.}
	\label{fig:AchievableRateSNR}
\end{figure}
Firstly, the system achievable rates under different SNRs are evaluated and the performance is compared between the proposed and conventional schemes for both HD and FD relayings.
The performance comparisons are shown in Fig. \ref{fig:AchievableRateSNR} where the SNR varies from -30 dB to 30 dB.
In the figure, the lines represent the rates obtained by Monte Carlo simulation while the markers denote the ergodic achievable rate lower bound computed by the closed-form expressions with statistical CSI (refer to as Theorem \ref{thm:DownlinkAchievableRate} and \ref{thm:UplinkAchievableRate}).
The comparison shows that the relative performance gap between Monte Carlo and the closed-form results is small, e.g., 1.76 bits/s/Hz for FD and 0.82 bits/s/Hz for HD at 0 dB of SNR, which implies that our closed-form ergodic achievable rate expression is a good predictor for the system performance.
In addition, Fig. \ref{fig:AchievableRateSNR} shows that the proposed scheme outperforms the conventional one in both high and low SNR regions, which verifies Corollary \ref{cry:AsymptoticResults}, where about 7.5 bits/s/Hz and 2.5 bits/s/Hz improvements are observed in the high SNR region for the FD and HD systems, respectively.
On top of that, the FD mode is shown exceeding the HD in system rates by about 13.6 bits/s/Hz in high SNRs with the proposed scheme, which verifies Remark \ref{rmk:Throughput}.

\begin{figure}[htbp]
	\centering
	\includegraphics[width=\linewidth]{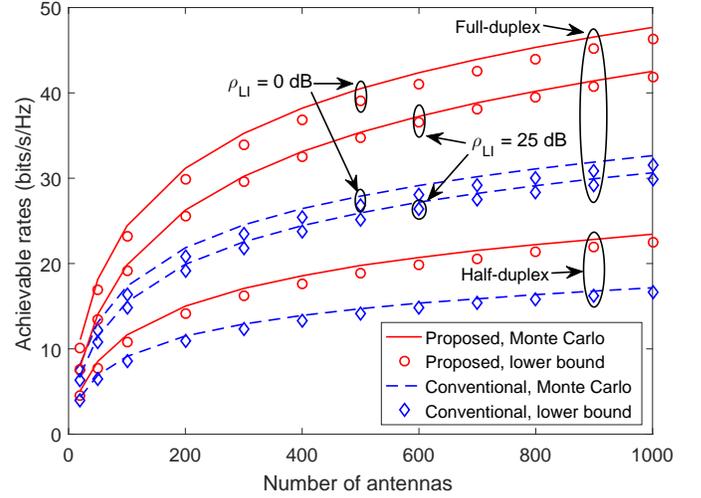}
	\caption{Comparisons between the proposed and conventional schemes versus the number of antennas equipped on RS.}
	\label{fig:AchievableRateNumAnt}
\end{figure}
Next, the performance of the massive MIMO system versus the growing number of RS antennas is depicted in Fig. \ref{fig:AchievableRateNumAnt}.
It is obvious that the rate gap between the proposed and conventional schemes increases as the number of RS antennas grows.
With more RS antennas, the source and destination channels are closer to be orthogonal to each other, and thus less interferences reside in the combined signal leading to improved performance of the proposed scheme.
As expected, the increase of the LI power degrades the FD performance and Fig. \ref{fig:AchievableRateNumAnt} shows the superiority of the proposed scheme in the FD system as the LI power equals to both as small as 0 dB and as large as 25 dB.
Further, even as few as 20 antennas are deployed on RS, the proposed scheme still outperforms the conventional one, which verifies that the proposed scheme is feasible and superior even in medium scale MIMO relaying systems.

\begin{figure}[htbp]
	\centering
	\includegraphics[width=\linewidth]{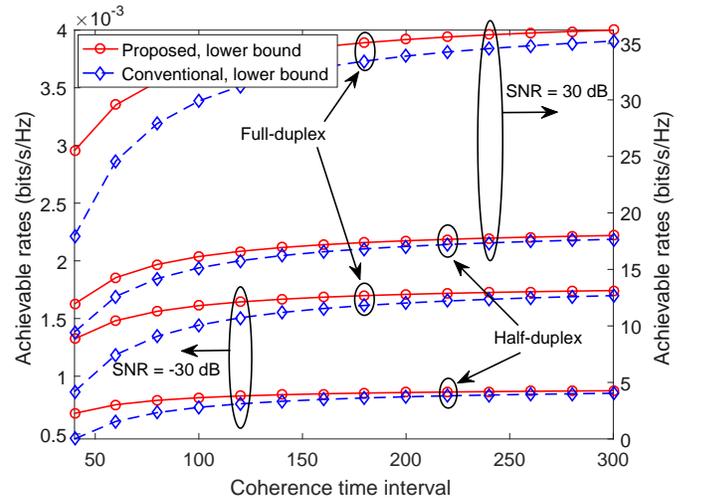}
	\caption{Impact of the coherence interval length on the performance of the proposed and conventional schemes.}
	\label{fig:AchievableRateCoTime}
\end{figure}
As discussed in Section \ref{sec:PowerAllocation}, further performance improvement of the proposed scheme can be obtained by balancing the tradeoff between pilot overhead and channel estimation accuracy where the estimation precision decreases due to the data interference as the percentage of data transmission time within a coherence interval increases.
Therefore, the length of the coherence interval and the SNR of the received pilots are two crucial system parameters that affect the channel estimation overhead and accuracy, respectively.
Here, the performance improvement of the proposed scheme is verified with respect to the length of coherence intervals.
According to Fig. \ref{fig:AchievableRateCoTime}, the achievable rate of the proposed scheme always outperforms the conventional one for coherence interval from 20 to 300 time slots at both high and low SNRs, which verifies the conclusion in Remark \ref{rmk:IndepentOfCoInterval}.
Nonetheless, the gap between the two schemes decreases with the increase of the interval length due to the reduction of relative pilot transmission overhead within a longer coherence interval.

\begin{figure}[htbp]
	\centering
	\includegraphics[width=\linewidth]{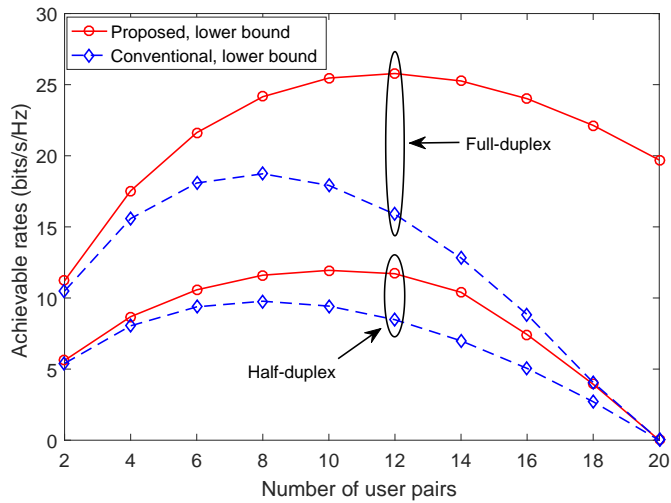}
	\caption{Comparisons between the proposed and conventional schemes versus the number of user pairs.}
	\label{fig:AchievableRateNumUser}
\end{figure}
In practical systems, the network operator would prefer to serve more user pairs to improve the overall performance of the entire system at long coherence intervals, especially when a large number of antennas exist\cite{Bjoernson2015,Bjoernson2016}.
Therefore, the performance of both pilot schemes by varying the number of user pairs is examined at a fixed coherence interval.
As shown in Fig. \ref{fig:AchievableRateNumUser}, the maximal rate is achieved when 12 user pairs are communicating simultaneously in the FD system by utilizing the proposed transmission scheme while only 8 user pairs are served with the conventional scheme whose sum rate rapidly deceases with the increase of user numbers.
Moreover, the proposed scheme outperforms the conventional one even with 2 user pairs served by the FD relaying system, which affirms the robustness of the proposed scheme when small number of users access to the network.
In the HD mode, similar performance comparisons are observed.
As expected, the growing number of user pairs also enlarges the rate gap between two pilot schemes.
Particularly, when the number of user pairs equals to the half length of coherence interval, i.e., $K=20$, all HD and conventional FD curves touch zero rates due to no time resource left for data transmission, except the FD overlay system still working in the high-throughput state.
Therefore, the comparisons reveals that both HD and FD relaying systems employing the pilot-data overlay scheme achieve higher system rates and serve more user pairs than those with the conventional pilot scheme.
Further, the FD overlay system emerges to be superior to all other systems in the extreme scenario.

\begin{figure}[htbp]
	\centering
	\includegraphics[width=\linewidth]{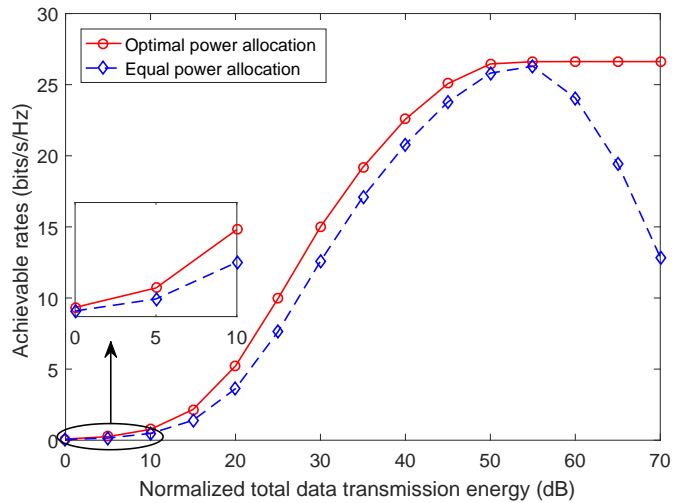}
	\caption{Performance comparison between two power allocation schemes in the FD overlay system.}
	\label{fig:AchievableRateSNRPowAlloc}
\end{figure}
\begin{figure}[htbp]
	\centering
	\includegraphics[width=\linewidth]{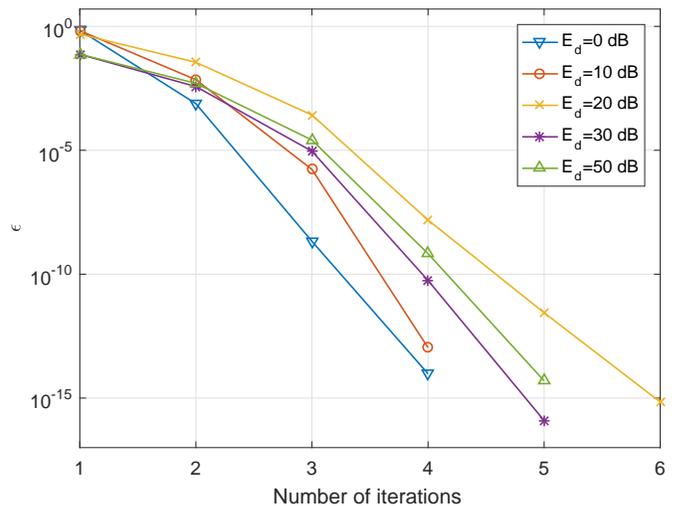}
	\caption{Convergence of the proposed SCA approach.}
	\label{fig:ConvergIteNumPowAlloc}
\end{figure}
Finally, the performance of the proposed power allocation approach is evaluated for the FD overlay relaying.
In the simulation, the pilot power is fixed to be 10 dB per user and the total data transmission power per coherence interval varies from -10 dB to 60 dB for both optimal and equal power allocations.
Additionally, $\rho_\mathrm{d}=K\rho_\mathrm{s}$ is set in the equal power allocation.
Fig. \ref{fig:AchievableRateSNRPowAlloc} shows the comparison of the system rates between the two allocation schemes while the convergence of the proposed approach at various cases are depicted in Fig. \ref{fig:ConvergIteNumPowAlloc}.
Fig. \ref{fig:AchievableRateSNRPowAlloc} shows that the optimal power allocation achieves better performance than the equal power allocation for both low and high transmission powers, especially at moderate power region, where the rate increment becomes more noticeable.
The rate of equal power allocation starts decreasing when the data transmission power increases to an extreme large value while the proposed scheme maintains a fixed high performance.
Such rate decrement of equal power allocation is due to more pilot contamination when data transmission power is large yet pilot power is fixed.
In contrast, the optimal power allocation scheme can adaptively adjust the data transmission power to control the interference between pilot and data transmissions and therefore always achieve the best performance.
In Fig. \ref{fig:ConvergIteNumPowAlloc}, the number of iterations is mostly below 4 if the relative error $\epsilon$ in Algorithm \ref{algtm:fd:SCA} is set to be $10^{-5}$, indicating the low complexity of the proposed approach.
Fig. \ref{fig:ConvergIteNumPowAlloc} shows that the proposed SCA approach always converges fast in solving the power allocation problem.

\section{Conclusion}\label{sec:Conclusion}
Channel estimation overhead is a critical limitation on improving performance of the multiuser massive MIMO systems.
Dealing this problem, this paper has proposed a pilot-data overlay transmission scheme in both HD and FD one-way multipair massive MIMO relaying systems.
The proposed scheme has exploited the orthogonality of source-relay and relay-destination channels with massive MIMO setups at the RS and redesigned the pilot and data transmission scheme to increase the effective data transmission period, which improves the achievable rate performance of the system in practice.
The asymptotic analysis in both low and high SNRs with infinite number RS antennas has been conducted in this paper and it has proven the superiority of the proposed scheme theoretically.
Finally, a power allocation problem algorithm based on SCA has been proposed for the FD system, which further increases the achievable rate of the relaying system.

\begin{appendix}
Note that the postfix $[\iota]$ is dropped for concision in following derivations.

\subsection{Preliminaries}
\begin{lemma}(The law of large numbers)\label{lm:LawLargeNumbers}
	\cite{cramer2004random} Let $\mathbf{p}$ and $\mathbf{q}$ are two mutually independent ${L}\times1$ random vectors consisting of i.i.d. $\CN(0, \sigma_p^2)$ and $\CN(0, \sigma_q^2)$ RVs, respectively.
	Then $\lim\limits_{{L}\rightarrow\infty}\mathbf{p}^\mathrm{H}\mathbf{p}/{L}\aseq\sigma_p^2$ and $\lim\limits_{{L}\rightarrow\infty}\mathbf{p}^\mathrm{H}\mathbf{q}/{L}\aseq0$.
\end{lemma}

\begin{lemma}\label{lm:FourthOrderMoment}
	\cite{Zhang2014} Let $\mathbf{x}$ and $\mathbf{y}$ be mutually independent $L\times 1$ random vectors following distributions $\CN(0, \sigma_x^2\mathbf{I}_L)$ and $\CN(0, \sigma_y^2\mathbf{I}_L)$, respectively. Then, $\E\{\|\mathbf{x}^\Hm\mathbf{x}\|^2\}=(L^2+L)\sigma_x^4$ and $\E\{\|\mathbf{x}^\Hm\mathbf{y}\|^2\}=L\sigma_x^2\sigma_y^2$.
\end{lemma}

\subsection{Proof of Proposition \ref{prp:SourceDataDetectionSB}}\label{apdx:prf:prp:SourceDataDetectionSB}
Without loss of generality, we focus on the data detection of the $k$th source user, where $k=1,2,\cdots,K$.
By highlighting the $k$th row of $\mathbf{\tilde{S}}^\mathrm{B}$ and normalizing its power, (\ref{eq:DataDetectionSB}) can be rewritten as follows:
\begin{equation}\label{eq:SourceDataDetectionSB}
	\begin{split}
		\mathbf{\hat{s}}_k^\mathrm{B} =& \frac{\mathbf{\tilde{s}}_k^\mathrm{B}}{\sqrt{\rho_\mathrm{s}}\left\|\mathbf{\hat{g}}_k\right\|^2}\\
		=&\frac{\sqrt{\rho_\mathrm{s}}\mathbf{\hat{g}}_{\mathrm{s}k}^\Hm\mathbf{G}_\mathrm{s}\mathbf{S}^\mathrm{B}
			+ \sqrt{K\rho_\mathrm{p}}\mathbf{\hat{g}}_{\mathrm{s}k}^\Hm\mathbf{G}_\mathrm{d}\bm{\Psi} 
			+ \mathbf{\hat{g}}_{\mathrm{s}k}^\Hm\mathbf{N}^\mathrm{B}}{\sqrt{\rho_\mathrm{s}}\left\|\mathbf{\hat{g}}_k\right\|^2}\\
		\overset{\mathrm{(a)}}{=} & \frac{\left(\mathbf{\hat{g}}_{\mathrm{s}k}^\Hm\mathbf{\hat{g}}_{\mathrm{s}k}+\mathbf{\hat{g}}_{\mathrm{s}k}^\Hm\bm{\varepsilon}_{\mathrm{s}k}\right)\mathbf{s}_k^\mathrm{B}}{\|\mathbf{\hat{g}}_{\mathrm{s}k}\|^2} + \frac{\sum_{i=1,i\neq k}^{K}\mathbf{\hat{g}}_{\mathrm{s}k}^\Hm\mathbf{g}_{\mathrm{s}i}\mathbf{s}^\mathrm{B}_i}{\left\|\mathbf{\hat{g}}_k\right\|^2}\\
		&+\frac{\sqrt{K\rho_\mathrm{p}}\sum_{i=1}^{K}\mathbf{\hat{g}}_{\mathrm{s}k}^\Hm\mathbf{g}_{\mathrm{d}i}\bm{\psi}_i}{\sqrt{\rho_\mathrm{s}}\left\|\mathbf{\hat{g}}_k\right\|^2}+\frac{\mathbf{\hat{g}}_{\mathrm{s}k}^\Hm\mathbf{N}^\mathrm{B}}{\sqrt{\rho_\mathrm{s}}\left\|\mathbf{\hat{g}}_k\right\|^2}\\
		\aseq & \mathbf{s}_k^\mathrm{B}
	\end{split}
\end{equation}
where $\mathbf{s}_i^\mathrm{B}$ and $\bm{\psi}_i$ are the $i$th rows of $\mathbf{S}^\mathrm{B}$ and $\bm{\Psi}$, respectively.
The derivation $\overset{\mathrm{(a)}}{=}$ attributes to the decomposition of source channels depicted by Eq. (\ref{eq:SourceChannelVectorDecompositionFirstInterval}).
Moreover, because $\mathbf{\hat{g}}_{\mathrm{s}k}^\Hm$ is independent of $\bm{\varepsilon}_{\mathrm{s}k}$, $\mathbf{g}_{\mathrm{s}i}$ ($i\neq k$) and $\mathbf{N}^\mathrm{B}$, by dividing both denominators and numerators by $M$ and applying Lemma \ref{lm:LawLargeNumbers}, the almost surely convergence in the last equality holds due to $M\rightarrow\infty$.
Therefore, $\mathbf{s}_k^\mathrm{B}$ can be exactly detected from $\mathbf{\tilde{s}}_k^\mathrm{B}$. 

\subsection{Proof of Theorem \ref{thm:DownlinkAchievableRate}}\label{apdx:prf:thm:DownlinkAchievableRate}
\begin{equation}\label{eq:ProofDownlinkE}
\E\left\{\mathbf{g}_{\mathrm{d}k}^\Hm\mathbf{\hat{g}}_{\mathrm{d}k}\right\}=\E\left\{\mathbf{\hat{g}}_{\mathrm{d}k}^\Hm\mathbf{\hat{g}}_{\mathrm{d}k}\right\}+\E\left\{\mathcal{E}_{\mathrm{d}k}^\Hm\mathbf{\hat{g}}_{\mathrm{d}k}\right\}=M\sigma_{\mathrm{d}k}^2
\end{equation}
where $\mathcal{E}_{\mathrm{d}k}$ is independent of $\mathbf{\hat{g}}_{\mathrm{d}k}$.
\begin{equation}\label{eq:ProofDownlinkVar}
\begin{split}
&\Var\left\{\mathbf{g}_{\mathrm{d}k}^\Hm\mathbf{\hat{g}}_{\mathrm{d}k}\right\}\\
=&\E\left\{|\mathbf{g}_{\mathrm{d}k}^\Hm\mathbf{\hat{g}}_{\mathrm{d}k}|^2\right\}-\left|\E\left\{\mathbf{g}_{\mathrm{d}k}^\Hm\mathbf{\hat{g}}_{\mathrm{d}k}\right\}\right|^2\\
=&\E\left\{|\mathbf{\hat{g}}_{\mathrm{d}k}^\Hm\mathbf{\hat{g}}_{\mathrm{d}k}|^2\right\}+\E\left\{|\mathcal{E}_{\mathrm{d}k}^\Hm\mathbf{\hat{g}}_{\mathrm{d}k}|^2\right\}-\left|\E\left\{\mathbf{g}_{\mathrm{d}k}^\Hm\mathbf{\hat{g}}_{\mathrm{d}k}\right\}\right|^2\\
\overset{\mathrm{(b)}}{=}&(M^2+M)\sigma_{\mathrm{d}k}^4+M\varepsilon_{\mathrm{d}k}^2\sigma_{\mathrm{d}k}^2-M^2\sigma_{\mathrm{d}k}^4\\
=&M\beta_{\mathrm{d}k}\sigma_{\mathrm{d}k}^2.
\end{split}
\end{equation}
where $\overset{\mathrm{(b)}}{=}$ exploits Lemma \ref{lm:FourthOrderMoment} and the independence between $\mathbf{\hat{g}}_{\mathrm{d}k}$ and $\mathcal{E}_{\mathrm{d}k}$.
MI defined by (\ref{eq:MIDownlink}) can be calculated as follows:
\begin{equation}\label{eq:ProofDownlinkMI}
	\mathrm{MI}_k^\mathrm{DL}=\rho_\mathrm{d}\alpha^2\sum_{i=1,i\neq k}^{K}\E\left\{\left|\mathbf{g}_{\mathrm{d}k}^\Hm\mathbf{\hat{g}}_{\mathrm{d}i}\right|^2\right\}
	\overset{\mathrm{(c)}}{=}M\rho_\mathrm{d}\alpha^2\beta_{\mathrm{d}k}\sum_{i=1,i\neq k}^{K}\sigma_{\mathrm{d}i}^2.
\end{equation}
where $\overset{\mathrm{(c)}}{=}$ exploits Lemma \ref{lm:FourthOrderMoment}.
By substituting (\ref{eq:ProofDownlinkE}) and (\ref{eq:ProofDownlinkMI}) into (\ref{eq:GammaDownlinkOrigin}), the desired result comes out.

\subsection{Proof of Theorem \ref{thm:UplinkAchievableRate}}\label{apdx:prf:thm:UplinkAchievableRate}
The expectation, variance and uplink MI terms in both (\ref{eq:GammaUplinkOriginB}) and (\ref{eq:GammaUplinkOriginC}) can be calculated by employing the same manipulations as Theorem \ref{thm:DownlinkAchievableRate} does.
The PI term is derived as
\begin{equation}
	\label{eq:ProofUplinkPI}
	\begin{split}
		\mathrm{PI}_k^\mathrm{UL} =&\rho_\mathrm{p}\tr\left(\E\left\{\mathbf{\hat{g}}_{\text{s}k}\mathbf{\hat{g}}_{\text{s}k}^\text{H}\mathbf{G}_{\text{d}}\mathbf{G}_{\text{d}}^\text{H}\right\}\right)\\
		\overset{\mathrm{(d)}}{=}&\rho_\mathrm{p}\tr\left(\sigma_{\text{s}k}^2\mathbf{I}_M\E\left\{\sum_{i=1}^{K}\mathbf{g}_{\text{d}i}\mathbf{g}_{\text{d}i}^\text{H}\right\}\right)\\
		=&M\rho_\mathrm{p}\sigma_{\text{s}k}^2\sum_{i=1}^{K}\beta_{\text{d}i}
	\end{split}
\end{equation}
where $\overset{\mathrm{(d)}}{=}$ exploits the independence between $\mathbf{\hat{g}}_{\text{s}k}$ and $\mathbf{G}_{\text{d}}$.
The LI term is obtained by
\begin{equation}
	\label{eq:ProofUplinkLI}
	\begin{split}
		\mathrm{LI}_k^\mathrm{UL}=&\rho_\mathrm{d}\alpha^2\tr\left(\E\left\{\mathbf{\hat{g}}_{\mathrm{s}k}\mathbf{\hat{g}}_{\mathrm{s}k}^\Hm\mathbf{G}_{\mathrm{LI}}\mathbf{\hat{G}}_\mathrm{d}\mathbf{\hat{G}}_\mathrm{d}^\Hm\mathbf{G}_{\mathrm{LI}}^\Hm\right\}\right)\\
		\overset{\mathrm{(e)}}{=}&\rho_\mathrm{d}\alpha^2\tr\left(\sigma_{\mathrm{s}k}^2\mathbf{I}_M\E\left\{\mathbf{G}_{\mathrm{LI}}^\Hm\mathbf{G}_{\mathrm{LI}}\right\}\E\left\{\mathbf{\hat{G}}_\mathrm{d}\mathbf{\hat{G}}_\mathrm{d}^\Hm\right\}\right)\\
		\overset{\mathrm{(f)}}{=}&\rho_\mathrm{d}\alpha^2\tr\left(\sigma_{\mathrm{s}k}^2\sum_{i=1}^{K}\sigma_{\mathrm{d}i}^2\mathbf{I}_M\E\left\{\mathbf{G}_{\mathrm{LI}}^\Hm\mathbf{G}_{\mathrm{LI}}\right\}\right)\\
		=&\rho_\mathrm{d}\alpha^2\tr\left(\sigma_{\mathrm{s}k}^2\sum_{i=1}^{K}\sigma_{\mathrm{d}i}^2M\beta_{\mathrm{LI}}\mathbf{I}_M\right)\\
		\overset{\mathrm{(g)}}{=}&M\rho_\mathrm{d}\beta_\mathrm{LI}^2\sigma_{\mathrm{s}k}^2
	\end{split}
\end{equation}
where $\overset{\mathrm{(e)}}{=}$ and $\overset{\mathrm{(f)}}{=}$ exploit the independences among $\mathbf{\hat{g}}_{\mathrm{s}k}$, $\mathbf{\hat{G}}_\mathrm{d}$ and $\mathbf{G}_{\mathrm{LI}}$ and the fact that $\tr(\mathbf{AB})=\tr(\mathbf{BA})$, and $\overset{\mathrm{(g)}}{=}$ takes $\alpha=\sqrt{1\big/\left(M\sum_{i=1}^{K}\sigma_{\mathrm{d}i}^2\right)}$.
Finally, the AWGN term is straightforwardly derived as $\mathrm{AN}_k^\mathrm{UL}=\sigma_{\mathrm{s}k}^2$.
By substituting above equations into (\ref{eq:GammaUplinkOriginB}) and (\ref{eq:GammaUplinkOriginC}), Theorem \ref{thm:UplinkAchievableRate} is obtained.

\subsection{Proof of Corollary \ref{cry:AsymptoticResults}}\label{apdx:prf:cry:AsymptoticResults}
\subsubsection{Half-duplex}
Here starts the proof with the high SNR case.
For the downlink, because $\lim\limits_{\rho\to\infty}\gamma_k^\mathrm{DL} = \lim\limits_{\rho\to\infty}\tilde{\gamma}_k^\mathrm{DL}$ and $T_d > \tilde{T}_d$, it is obvious that
\begin{equation}\label{eq:hd:prf:HighSNRDownlink}
	\lim\limits_{\rho\to\infty} \mathcal{R}_k^\mathrm{DL} > \lim\limits_{\rho\to\infty}\mathcal{\tilde{R}}_k^\mathrm{DL}.
\end{equation}
For the uplink, the difference of the rates between the proposed and conventional schemes is evaluated as:
\begin{equation}\label{eq:hd:prf:HighSNRUplink1}
	\begin{split}
		&\lim_{\rho\to\infty}\left(\mathcal{R}_k^\mathrm{UL} - \mathcal{\tilde{R}}_k^\mathrm{UL}\right)\\
		= & K\log_2\left(1+\lim_{\rho\to\infty}\gamma_k^\mathrm{B}\right) + (T_d-K)\log_2\left(1+\lim_{\rho\to\infty}\gamma_k^\mathrm{C}\right)\\
		& - \tilde{T}_d\log_2\left(1+\lim_{\rho\to\infty}\tilde{\gamma}_k^\mathrm{UL}\right)\\
		= &\frac{K}{2}\log_2\left(1+\frac{M\beta_{\text{s}k}}{\sum_{i=1}^{K}\beta_{\text{s}i}+\sum_{i=1}^{K}\beta_{\text{d}k}}\right)^2\\
		&-\frac{K}{2}\log_2\left(1+\frac{M\beta_{\text{s}k}}{\sum_{i=1}^{K}\beta_{\text{s}i}}\right)
	\end{split}
\end{equation}
where $T_d=(T_c-K)/2$ and $\tilde{T}_d=(T_c-2K)/2$.
Hence, 
\begin{equation}\label{eq:hd:prf:HighSNRUplink}
	\lim_{\rho\to\infty}\mathcal{R}_k^\mathrm{UL} \geq \lim_{\rho\to\infty}\mathcal{\tilde{R}}_k^\mathrm{UL}
\end{equation}
holds, if and only if
\begin{equation}\label{eq:hd:prf:HighSNRUplink2}
\left(1+\frac{M\beta_{\text{s}k}}{\sum_{i=1}^{K}\beta_{\text{s}i}+\sum_{i=1}^{K}\beta_{\text{d}k}}\right)^2\geq 1+\frac{M\beta_{\text{s}k}}{\sum_{i=1}^{K}\beta_{\text{s}i}}
\end{equation}
which is equivalent to
\begin{equation}\label{eq:hd:prf:HighSNRUplink3}
\left(\sum_{i=1}^{K}\beta_{\text{s}i}\right)^2+M\beta_{\text{s}k}\sum_{i=1}^{K}\beta_{\text{s}i} \geq \left(\sum_{i=1}^{K}\beta_{\text{d}i}\right)^2
\end{equation}
where the left-hand-side term is $M$-dependent while the right-hand-side term is a fixed finite value.
In massive MIMO systems, it is always assumed that $M\gg K$ and therefore (\ref{eq:hd:prf:HighSNRUplink3}) holds with finite $\beta_{\text{s}i}$ and $\beta_{\text{d}i}$.
Hence, (\ref{eq:hd:prf:HighSNRUplink}) holds, with respect to any $k$ within $1$ to $K$.
Therefore, the proposed scheme outperforms the conventional one for high SNRs in half-duplex communications by combining (\ref{eq:hd:prf:HighSNRDownlink}) and (\ref{eq:hd:prf:HighSNRUplink}).

On the other hand, regarding low SNRs, it is obvious that 
\begin{align}
	\lim_{\rho\to 0}\gamma_k^\mathrm{UL} =& \lim_{\rho\to 0}\tilde{\gamma}_k^\mathrm{UL} \\
	\lim_{\rho\to 0}\gamma_k^\mathrm{DL} =& \lim_{\rho\to 0}\tilde{\gamma}_k^\mathrm{DL}.
\end{align}
Therefore, the achievable rate of the proposed scheme is greater than those of the conventional one due to the multiplication factor $T_d > \tilde{T}_d$.

\subsubsection{Full-duplex}
For the FD, $T_d=T_c-K$ and $\tilde{T}_d=T_c-2K$.
With similar manipulations as in HD, it is straightforward to prove that the proposed scheme always outperforms the conventional one for both low and high SNRs.

\end{appendix}


\begin{thebibliography}{10}
	\providecommand{\url}[1]{#1}
	\csname url@samestyle\endcsname
	\providecommand{\newblock}{\relax}
	\providecommand{\bibinfo}[2]{#2}
	\providecommand{\BIBentrySTDinterwordspacing}{\spaceskip=0pt\relax}
	\providecommand{\BIBentryALTinterwordstretchfactor}{4}
	\providecommand{\BIBentryALTinterwordspacing}{\spaceskip=\fontdimen2\font plus
		\BIBentryALTinterwordstretchfactor\fontdimen3\font minus
		\fontdimen4\font\relax}
	\providecommand{\BIBforeignlanguage}[2]{{%
			\expandafter\ifx\csname l@#1\endcsname\relax
			\typeout{** WARNING: IEEEtran.bst: No hyphenation pattern has been}%
			\typeout{** loaded for the language `#1'. Using the pattern for}%
			\typeout{** the default language instead.}%
			\else
			\language=\csname l@#1\endcsname
			\fi
			#2}}
	\providecommand{\BIBdecl}{\relax}
	\BIBdecl
	
	\bibitem{Marzetta2010}
	T.~L. Marzetta, ``{Noncooperative cellular wireless with unlimited numbers of
		base station antennas},'' \emph{IEEE Trans. Wireless Commun.}, vol.~9,
	no.~11, pp. 3590--3600, Nov. 2010.
	
	\bibitem{Hieving2013}
	F.~Rusek, D.~Persson, B.~K. Lau, E.~G. Larsson, T.~L. Marzetta, O.~Edfors, and
	F.~Tufvesson, ``{Scaling up MIMO: Opportunities and challenges with very
		large arrays},'' \emph{IEEE Signal Process. Mag.}, vol.~30, no.~1, pp.
	40--60, Jan. 2013.
	
	\bibitem{Hoydis2013}
	J.~Hoydis, S.~{Ten Brink}, and M.~Debbah, ``{Massive MIMO in the UL/DL of
		cellular networks: How many antennas do we need?}'' \emph{IEEE J. Sel. Areas
		Commun.}, vol.~31, no.~2, pp. 160--171, Feb. 2013.
	
	\bibitem{Larsson2014Massive}
	E.~G. Larsson, O.~Edfors, F.~Tufvesson, and T.~L. Marzetta, ``{Massive MIMO for
		next generation wireless systems},'' \emph{IEEE Commun. Mag.}, vol.~30,
	no.~1, pp. 186--195, Feb. 2014.
	
	\bibitem{Chin2014}
	W.~Chin, Z.~Fan, and R.~Haines, ``{Emerging technologies and research
		challenges for 5G wireless networks},'' \emph{IEEE Wireless Commun.},
	vol.~21, no.~2, pp. 106--112, Apr. 2014.
	
	\bibitem{Marzetta2006}
	T.~L. Marzetta, ``{How much training is required for multiuser MIMO?}'' in
	\emph{Proc. Asilomar Conf. Signals, Syst., Comput.}, Oct. 2006, pp. 359--363.
	
	\bibitem{Jose2011}
	J.~Jose, A.~Ashikhmin, T.~L. Marzetta, and S.~Vishwanath, ``{Pilot
		contamination and precoding in multi-cell TDD systems},'' \emph{IEEE Trans.
		Wireless Commun.}, vol.~10, no.~8, pp. 2640--2651, Aug. 2011.
	
	\bibitem{Zhang2014}
	H.~Zhang, X.~Zheng, W.~Xu, and X.~You, ``{On massive MIMO performance with
		semi-orthogonal pilot-assisted channel estimation},'' \emph{EURASIP J.
		Wireless Commun. and Netw.}, vol. 2014, pp. 1--14, Dec. 2014.
	
	\bibitem{Zheng2014}
	X.~Zheng, H.~Zhang, W.~Xu, and X.~You, ``{Semi-orthogonal pilot design for
		massive MIMO systems using successive interference cancellation},'' in
	\emph{Proc. IEEE GLOBECOM}, Dec. 2014, pp. 3719--3724.
	
	\bibitem{You2015}
	L.~You, X.~Gao, X.-G. Xia, N.~Ma, and Y.~Peng, ``{Pilot reuse for massive MIMO
		transmission over spatially correlated rayleigh fading channels},''
	\emph{IEEE Trans. Wireless Commun.}, vol. 1276, no.~6, pp. 1--15, Jun. 2015.
	
	\bibitem{Fernandes2013}
	F.~Fernandes, A.~Ashikhmin, and T.~Marzetta, ``{Inter-cell interference in
		noncooperative TDD large scale antenna systems},'' \emph{IEEE J. Sel. Areas
		Commun.}, vol.~31, no.~2, pp. 192--201, Feb. 2013.
	
	\bibitem{Jin2014}
	S.~Jin, X.~Wang, Z.~Li, and K.-K. Wong, ``{Zero-forcing beamforming in massive
		MIMO systems with time-shifted pilots},'' in \emph{Proc. IEEE ICC}, Jun.
	2014, pp. 4801--4806.
	
	\bibitem{DohlerLi2010}
	M.~Dohler and Y.~Li, \emph{{Cooperative Communications: Hardware, Channel and
			PHY}}.\hskip 1em plus 0.5em minus 0.4em\relax John Wiley \& Sons, Jan. 2010.
	
	\bibitem{Ngo2013}
	H.~Q. Ngo and E.~G. Larsson, ``{Spectral efficiency of the multipair two-way
		relay channel with massive arrays},'' in \emph{Proc. Asilomar Conf. Signals,
		Syst., Comput.}, Nov. 2013, pp. 275--279.
	
	\bibitem{NgoLarssonMemberEtAl2013}
	H.~Q. Ngo, E.~G. Larsson, S.~Member, and S.~Member, ``{Large-scale multipair
		two-way relay networks with distributed AF beamforming large-scale multipair
		two-way relay networks with distributed AF beamforming},'' \emph{IEEE Commun.
		Lett.}, vol.~17, no.~12, pp. 2288--2291, Dec. 2013.
	
	\bibitem{Suraweera2013}
	H.~A. Suraweera, H.~Q. Ngo, T.~Q. Duong, C.~Yuen, and E.~G. Larsson,
	``{Multi-pair amplify-and-forward relaying with very large antenna arrays},''
	in \emph{Proc. IEEE ICC}, Jun. 2013, pp. 4635--4640.
	
	\bibitem{Dai2015}
	\BIBentryALTinterwordspacing
	Y.~Dai and X.~Dong, ``{Power allocation for multi-pair massive MIMO two-way AF
		relaying with robust linear processing},'' \emph{arXiv preprint
		arXiv:1508.06656}, pp. 1--14, Aug. 2015. [Online]. Available:
	\url{http://arxiv.org/abs/1508.06656}
	\BIBentrySTDinterwordspacing
	
	\bibitem{Ngo2014}
	H.~Q. Ngo, H.~A. Suraweera, M.~Matthaiou, and E.~G. Larsson, ``{Multipair
		full-duplex relaying with massive arrays and linear processing},'' \emph{IEEE
		J. Sel. Areas Commun.}, vol.~32, no.~9, pp. 1721--1737, Sept. 2014.
	
	\bibitem{Ngo2014a}
	H.~Q. Ngo, H.~a. Suraweera, M.~Matthaiou, and E.~G. Larsson, ``{Multipair
		massive MIMO full-duplex relaying with MRC/MRT processing},'' in \emph{Proc.
		IEEE ICC}, Jun. 2014, pp. 4818--4824.
	
	\bibitem{Rodriguez2014}
	L.~J. Rodriguez, N.~H. Tran, and T.~Le-Ngoc, ``{Optimal power allocation and
		capacity of full-duplex AF relaying under residual self-interference},''
	\emph{IEEE Wireless Commun. Lett.}, vol.~3, no.~2, pp. 233--236, Apr. 2014.
	
	\bibitem{Riihonen2011}
	T.~Riihonen, S.~Werner, and R.~Wichman, ``{Mitigation of loopback
		self-interference in full-duplex MIMO relays},'' \emph{IEEE Trans. Signal
		Process.}, vol.~59, no.~12, pp. 5983--5993, Dec. 2011.
	
	\bibitem{Bharadia2014}
	D.~Bharadia and S.~Katti, ``{Full duplex MIMO radios},'' in \emph{Proc. USENIX
		NSDI}, Apr. 2014, pp. 359--372.
	
	\bibitem{cui2014multi}
	H.~Cui, L.~Song, and B.~Jiao, ``{Multi-pair two-way amplify-and-forward
		relaying with very large number of relay antennas},'' \emph{IEEE Trans.
		Wireless Commun.}, vol.~13, no.~5, pp. 2636--2645, May 2014.
	
	\bibitem{cramer2004random}
	H.~Cram{\'e}r, \emph{{Random Variables and Probability Distributions}}.\hskip
	1em plus 0.5em minus 0.4em\relax Cambridge University Press, Jun. 2004,
	vol.~36.
	
	\bibitem{Moon2000}
	T.~K. Moon and W.~C. Stirling, \emph{{Mathematical Methods and Algorithms for
			Signal Processing}}.\hskip 1em plus 0.5em minus 0.4em\relax Prentice Hall New
	York, 2000, vol.~1.
	
	\bibitem{Hasna2004}
	M.~O. Hasna and M.-S. Alouini, ``{Optimal power allocation for relayed
		transmissions over Rayleigh-fading channels},'' \emph{IEEE Trans. Wireless
		Commun.}, vol.~3, no.~6, pp. 1999--2004, Nov. 2004.
	
	\bibitem{Chiang2007}
	M.~Chiang, C.~W. Tan, D.~P. Palomar, D.~O'Neill, and D.~Julian, ``{Power
		control by geometric programming},'' \emph{IEEE Trans. Wireless Commun.},
	vol.~6, no.~7, pp. 2640--2651, Jul. 2007.
	
	\bibitem{Boyd2004}
	S.~Boyd and L.~Vandenberghe, \emph{{Convex Optimization}}.\hskip 1em plus 0.5em
	minus 0.4em\relax Cambridge University Press, Mar. 2004.
	
	\bibitem{Bjoernson2015}
	E.~Bj{\"{o}}rnson, L.~Sanguinetti, J.~Hoydis, and M.~Debbah, ``{Optimal design
		of energy-efficient multi-user MIMO systems: Is massive MIMO the answer?}''
	\emph{IEEE Trans. on Wireless Commun.}, vol.~14, no.~6, pp. 3059 -- 3075,
	Jun. 2015.
	
	\bibitem{Bjoernson2016}
	E.~Bj{\"{o}}rnson, E.~G. Larsson, and M.~Debbah, ``{Massive MIMO for maximal
		spectral efficiency: How many users and pilots should be allocated?}''
	\emph{IEEE Trans. on Wireless Commun.}, vol.~15, no.~2, pp. 1293 -- 1308,
	Feb. 2016.
	
\end{thebibliography}


\end{document}